\newcommand{\Mpc}{\, \rm{Mpc} }
\newcommand{\cMpc}{\, \rm{cMpc} }

\documentclass[fleqn,usenatbib]{mnras}

\usepackage{newtxtext,newtxmath}

\usepackage[T1]{fontenc}

\DeclareRobustCommand{\VAN}[3]{#2}
\let\VANthebibliography\thebibliography
\def\thebibliography{\DeclareRobustCommand{\VAN}[3]{##3}\VANthebibliography}

\usepackage{graphicx}	
\usepackage{amsmath}	

\title[X-ray heating in different 21-cm simulation codes]{How X-rays heat the IGM in different 21-cm simulation codes: a comparison between \textsc{Licorice} and \textsc{Beorn}  }

\author[Romain Meriot et al.]{
Romain Meriot,$^{1,2}$\thanks{E-mail: rmeriot@mpifr-bonn.mpg.de}
Jonathan R.~Pritchard,$^{1,2}$
and Timothée Schaeffer$^{3}$
\\
$^{1}$Max Planck Institute for Radio Astronomy, Auf dem Hügel 69, 53121 Bonn, Germany\\
$^{2}$Imperial College London, Prince Consort Road, London, SW7 2AZ, UK\\
$^{3}$Department of Astrophysics, University of Zurich,
Winterthurerstrasse 190, 8057 Zurich, Switzerland.
}

\date{Accepted XXX. Received YYY; in original form ZZZ}

\pubyear{\the\year{}}

\begin{document}
\label{firstpage}
\pagerange{\pageref{firstpage}--\pageref{lastpage}}
\maketitle

\begin{abstract}

Any interpretation of the 21-cm signal of neutral hydrogen using Bayesian inference methods can only be as accurate as the underlying simulation code {used to model the state of the intergalactic medium (IGM)}. 3D radiative transfer (RT) simulation codes may capture complex physics, but are computationally expensive and, {therefore}, faster, more approximate codes have been developed. To improve our understanding of the convergence of simulation codes in the 21-cm science community, we present a comparison of the X-ray heating of the IGM modelled in \textsc{Licorice}, a 3D RT simulation code, and \textsc{Beorn}, a 1D RT code. We use \textsc{Beorn} to process sources extracted from \textsc{Licorice} simulations, using the same physics of the sources, in order to obtain two versions of the temperature of the IGM heated by X-rays. We observe a good agreement between the luminosity fields, { mean temperatures, and global 21-cm signal } of the two setups, but discrepancies in the distribution of temperature and 21-cm signal, which result in a $\sim 30\%$ difference in the 21-cm power spectrum. We attempt to isolate the approximations that lead to these differences and find that {common approximations used in 1D RT} codes produce effects of that magnitude. Using an emulator of the \textsc{Licorice} power spectra in an MCMC pipeline, we translate these differences between power spectra into differences between posterior distributions over the astrophysical parameters.  We observe a typical bias between 1D posteriors of $\gtrsim 1 \sigma$ {(with a noise level corresponding to 100h of SKA observations)}. 

\end{abstract}




\section{Introduction}

The 21-cm signal of neutral hydrogen is set to be a treasure trove of information regarding the first billion years of the Universe, from the birth of the first luminous objects during the Cosmic Dawn (CD) to the  end of the Epoch of Reionisation (EoR) \citep[e.g.][]{Furlanetto2006a, Pritchard2012}. The precise observation of the signal by the Square Kilometre Array (SKA) should give an unprecedented window into the early Universe, and SKA precursors such as HERA\footnote{Hydrogen epoch of Reionization Array.}, LOFAR\footnote{Low-Frequency Array.}, Nenufar\footnote{New Extension in Nançay Upgrading LOFAR.}, MWA\footnote{Murchison Widefield Array.},  aim to measure the power spectrum of the fluctuations of the signal.

Detecting the signal is difficult: foregrounds are approximately four orders of magnitude more intense than the signal and must either be subtracted or avoided. The former strategy (foreground removal, used in e.g. the LOFAR processing pipeline) requires a careful modelling of the foregrounds as not to leave intense residuals or subtract part of the signal alongside the foregrounds, and the latter (foreground avoidance, adopted by e.g. HERA) focuses the observations on a relatively foreground-free "EoR-window" of Fourier space, and by doing so limits the amount of observable information. In any case, the instruments are vulnerable to many sources of systematics caused for instance by a slightly imperfect calibration or by contamination by man-made radio-frequency interference. The growth in understanding of the instruments and of numerous causes of systematics has led to a steady improvement of the upper limits on the 21-cm power spectrum during the EoR and CD, and most recent limits slightly constrain standard models of the signal \citep{Abdurashidova2022, heraphaseii, Mertens2025}. 

As a complement to interferometric observations, single-dish experiments such as EDGES \citep{Murray2026}, SARAS \citep{Nambissan2021a}, REACH \citep{LeraAcedo} wish to measure the sky-averaged global 21-cm signal and must face similar obstacles. A tentative detection by EDGES was claimed in \cite{Bowmana}, suggesting that the signal had a deep $\sim -500$ mK absorption trough at $z\sim20$. This observation was rejected by SARAS3 with a $95 \%$ confidence level \citep{Nambissan2021a, Singh2022},  and recent Bayesian reinterpretations of the EDGES data favour a result dominated by systematics \citep{Cang2024, Sims2020}. A new generation of instruments is underway, with efforts to place single or multi-antennas instruments on the Lunar Farside, which for the time being remains extremely radio-quiet \citep[e.g.][]{burns21, Brinkerink, Chen2024, Burns21a, Polidan2024}.

While the observation of the signal is a massive instrumental challenge, interpreting current and future measurements requires inference pipelines relying on complex simulation codes. Inference methods in the context of 21-cm cosmology have been the focus of  many projects over the past few years with the development of a family of inference techniques named Simulation-Based Inference \citep[see e.g.][]{Papamakarios, Alsing2018, Alsing2019, Zhao2022, Prelogovic2023, Cooper2025}. Their performances have been evaluated and compared in recent works \citep[e.g.][]{Meriot2025, Prelogovic2023}, and the efforts of the community to study summary statistics and extract as much astrophysical information as possible from observations are ongoing \citep[e.g.][]{Prelogovic2024, Cerardi2025}.

Simulating the 21-cm signal has been an ongoing effort over the past few decades and the community has developed several codes spanning a wide range of complexity and cost, which can be broadly sorted into three categories. Semi-numerical codes such as \textsc{21cmFAST} \citep[e.g.][]{Mesinger2010, Sobacchi2014, Park2018},   \textsc{21cmSPACE} \citep[][and references therein]{Sikder2026, Cohen2020} are fast, efficient codes which rely on many approximations, the most crucial of which likely being the use of simple "photon budget" arguments to compute the ionisation state of the IGM, thus bypassing the simulation of radiative transfer.  1D radiative transfer codes such as \textsc{Grizzly} \citep{Ghara2018}, \textsc{Beorn} \citep{Schaeffer2023}, Bears \citep{Thomas2009, Krause2018}, paint pre-computed temperature profiles around sources found in independently run N-body simulations. Finally, some 3D radiative transfer codes are also used by the EoR community, such as  C2-ray \citep{Mellema2006, Hirling2024}, Emma \citep{Aubert2015}, \textsc{Licorice} \citep{Meriot2024, Semelin2017, Semelin2007}. This paper makes use of two state-of-the-art codes: \textsc{Beorn} and \textsc{Licorice}.

The most recent version of \textsc{Licorice}, as well as the Lyman-alpha semi-numerical 1D radiative transfer code \textsc{SPINTER} \citep{Semelin2023a}, were used to create Loreli II, a database of nearly 10 000 simulations run at $256^3$ resolution \citep{Meriot2025, Meriot2024}. These simulations are samples of a reasonable prior distribution over the astrophysical parameters, and this sampling is dense enough to allow for correct inference results on mock data affected by a realistic noise realisation corresponding to 100 hours of SKA observations. However, \cite{Prelogovic2023, Meriot2025} suggest that the density of Loreli II in parameter space is insufficient to deal with lower noise levels. Loreli II required a few million CPU-hours, and a prospective Loreli III containing $\sim 10^5$ simulations would be a very costly project and the gains may not materialize until the SKA can deliver hundreds or thousands of hours of systematics-free data, which will not happen before the next decade.

To add to this, the community is using a variety of cheaper and faster codes to simulate the signal. If the physical approximations made to speed up these codes still allow for an accurate modelling of the Universe, {it becomes harder } to justify the cost of large databases of 3D radiative-transfer simulations. However, the consequences of approximations and design choices and the variance between simulation codes that follows are still relatively poorly understood in the 21-cm community. 

Two data challenges were recently organised by the SKA organisation 
to incentivize the development and comparisons of independent pipelines. SKA data challenge 3a (SDC3a) focused on the extraction of the signal from raw observations \citep{Bonaldi2025}, while SDC3b was an opportunity for many teams to perform inference on the same mock observations to obtain the mean neutral fraction of the Universe at several redshifts. The preliminary results\footnote{https://sdc3.skao.int/challenges/inference/results} of the SDC3b unfortunately suggest that the variance between different models, inference methods, and preprocessing pipelines, may be considerable.

Some thorough comparisons between simulation codes have been made in the past \citep{Hutter2018, Ghara2018, Majumdar2014, Zahn2010}, and the differences between the 21-cm power spectra of 3D RT simulations and faster codes seem to be significant, of the order of $\sim 20\%$ at moderate redshifts and  relatively low scales, where the thermal noise will be small. Existing works have tended to focus on the differences in the ionisation field and assume a saturated spin temperature, thus ignoring the differences in the X-ray heating, and to rely on a very small number of 3D RT simulations {(typically one or two)}, as the computation cost prohibited the exploration of parameter space with such expensive models.  

In this paper, we run the 1D radiative transfer code \textsc{Beorn} on several \textsc{Licorice} simulations to explore the differences in the effect of X-rays on the IGM{, since X-ray heating has not been considered in previous comparison papers}. {Although we compare two specific codes, we expect our results to highlight both the challenges of verifying the agreement of the different simulation codes used by the 21-cm community and the impact of certain approximations that commonly exist within those codes. }

This paper is structured as follows. In Section \ref{sec:lico_Beorn}, we describe the two simulation codes and our methodology for matching the source populations. In Section \ref{sec:results}, we compare the resulting temperature and 21-cm brightness-temperature fields, as well as their power spectra. We investigate the impact of the observed differences on astrophysical parameter inference using a Bayesian framework. Finally, we summarise our conclusions and discuss future perspectives in Section \ref{sec:concl}. A number of detailed numerical points are deferred to the Appendix.

\section{\textsc{Licorice} and \textsc{Beorn} }\label{sec:lico_Beorn}

\subsection{\textsc{Licorice}: a hydro-radiative cosmological code} 

Here we give some details about the  \textsc{Licorice} simulation code. Throughout this paper, the focus will be on the X-rays; for a complete description of the database and the code, see  \cite{Meriot2025} and references therein. \textsc{Licorice} is an EoR simulation code that couples smoothed-particles-hydrodynamics (SPH) with Monte-carlo ray-tracing 3D radiative transfer.    

SPH evaluates the density of gas particles, and dense gas particles form stars through one of two channels: either a given particle is denser than a threshold (set at 200 times the average density of the universe) and is considered fully collapsed and star forming, or it is below this threshold and we estimate unresolved star formation using a subgrid model described in \cite{Meriot2024}. This model relies on computing the halo mass function of a particle depending on its density and volume, and yields an estimate of the fraction of the gas mass that is collapsing and fueling star formation in halos more massive than a parameter $M_{min}$. A fully or partially collapsed gas particle then converts a fraction of its gas mass into stellar mass at each time step over a timescale $\tau_{SF}$. 

We then ascribe a UV luminosity $L_{UV}$ to the star forming particle taking into account the stellar lifetimes of the stellar population we model, a process described in more detail in \cite{Baek2010a}. Note that we do not explicitly use the common assumption that the UV luminosity of a galaxy is proportional to its star formation rate \citep[see e.g.][]{Park2018}, although we find this assumption to broadly hold on average in our simulations.

The X-ray luminosity $L_{X}$ of the particle is derived from its UV luminosity through
\begin{equation}
    L_{X} = 0.00192 f_X L_{UV} ,
\end{equation}

\noindent where $f_X$ is a free parameter and $0.00192$ is a fiducial normalisation factor. We include both "soft" ($E < 2 \, \rm{keV}$) and hard ($E > 2 \, \rm{keV}$) X-rays, emitted by {Quasi-stellar objects (QSO) and X-ray Binaries (XRB)} respectively, and we parametrize the energy fraction contained in hard X-rays with {a hard-to-soft ratio parameter} $r_{H/S}$. 
Gas particles with non-zero stellar masses are considered sources and send monochromatic photon packets through the simulated box. The frequencies of the UV and X-ray packets are drawn from their respective spectra. The photon packets travel in the simulation volume, and deposit their energy in simulated cells they encounter. The ionisation and temperature of the particles contained in these cells are then updated. 

Since we are interested in the effect of X-rays, the relevant quantity is the temperature of the neutral gas $T_{HI}$, updated following this equation: 
\begin{equation}\label{eq:tphi}
        \frac{dT_{HI}}{dt} = \frac{2}{3 k_B n_{HI}}\left[ - \frac{3}{2}k_B T_{HI}\frac{dn_{HI}}{dt} + \Lambda_{HI} + K_{HI} \right],
\end{equation}

\noindent where $n_{HI}$ is the number density of the hydrogen atoms, $\Lambda_{HI}$ is the X-ray heating rate, and $K_{HI}$ is the energy change of the particle as it adiabatically expands or contracts, as well as the effect of the SPH viscosity \citep[e.g.][]{Springel2005}. 

As an X-ray packet of $N_\gamma$ photons of frequency $\nu$ passes through a cell over a length $dl$ during a time $dt$, the optical depth $d\tau_{HI}$ is computed according to:
\begin{equation}
    d\tau_{HI} = n_{HI, cell} \sigma_{HI}(\nu)dl.
\end{equation}

For a particle in that cell, we compute the heating rate  $ \Lambda_{HI}$ following 
\begin{equation}\label{eq:heat_rate_lico}
    \Lambda_{HI} = N_\gamma(1-e^ {-d\tau_{HI}} ) h_P ( \nu - \nu_{HI}) f_{heat}(x_{ion})\chi /dt  
\end{equation}

\noindent where {$\chi$ is the ratio of the mass of a gas particle and the mass contained in the cell}, $h_P$ is the Planck constant, $f_{heat}$ is the fraction of the X-ray energy deposited as heat, computed following the fit of \cite{Shull1985}:  
\begin{equation}\label{eq:fheat}
    f_{heat}(x_{ion}) = 0.9971  \left[ 1- (1-x_{ion}^{0.2663}) \right]^{1.3163},
\end{equation}

\noindent where $x_{ion}$ is the local ionised fraction of a cell. 

{In the following, we consider simulations of size $(300 \cMpc/h)^3$ that ran with $256^3$ particles (half of which are dark matter and half gas) from redshift 100 down to redshift 8. We choose to stop the simulations before Reionization as we wish to focus on the heating by X-rays and ignore the UV photons. We use initial conditions computed with MUSIC \citep{Hahn2011} using the cosmological parameters of Planck \citep{Aghanim2020}. }



\subsection{\textsc{Beorn}: a 1D RT simulation code}

\textsc{Beorn} is a 1D radiative transfer code that "paints" pre-computed temperature, ionisation, and Lyman-alpha profiles on a 3D grid, around the sources, similar in approach to other 1D RT codes such as {BEARS \citep{Thomas2009}} or \textsc{Grizzly} \citep{Ghara2018}. 
We summarise the relevant elements of the code here, and the details of the code can be found in \cite{Schaeffer2023, Schneider}.

\textsc{Beorn} takes as input a halo accretion history and computes the corresponding 1D temperature profile and its evolution, under certain assumptions.  In particular, it assumes an analytical relation between the masses of the halo and their star formation rate. The X-ray emissivity is then assumed to be proportional to the star formation rates.
Indeed, the X-ray flux profiles $\rho_X$ of \textsc{Beorn} are computed following:
\begin{equation}
    \rho_X(r,z) = \frac{1}{4 \pi r^2} \sum_i h_P f_{heat} \int_{\nu_{HI}} d\nu ( \nu - \nu_{HI}) e^{\tau_{\nu'}}  \dot{M_*}(z') \epsilon_X(\nu').
\end{equation}

In this equation, $\dot{M_*}$ is the star formation rate of the source, $\nu_{HI} \approx 3.29 \times 10^{15}$ Hz is the frequency of hydrogen ionisation, and
\begin{equation}
    \epsilon_X(\nu) = \left(3.4 \times 10^{40} \rm{erg.s^{-1}}\right) f_X \frac{I_X(\nu)}{h_P \nu},
\end{equation}
 \noindent where the spectrum $I_X$ follows a power law of index $n_X$, between frequencies parametrized by $E_{min}$ and $E_{max}$, and only takes into account the `soft' X-rays emitted by QSOs. {$\tau$ is the optical depth of the IGM, computed using the mean value of the density of the universe.}
 
These X-ray profiles appear in the differential equation used to compute the temperature profiles $\rho_T$: 
\begin{equation}\label{eq:rho_heat}
    \frac{3}{2} \frac{d\rho_T}{dz}(r,z) = \frac{3}{(1+z)} \rho_T(r,z) - \frac{1}{k_B (1+z) H(z)} \rho_X(r,z).
\end{equation}

The initial ingredient required to compute a profile is a halo accretion history. In the public version of the code presented in \cite{Schaeffer2023}, the masses of halos are assumed to follow
\begin{equation}
    M_h(z) = M_h(z_0)\rm{exp}[-\alpha (z-z_0)],  
\end{equation}

\noindent with parameters $\alpha$ and $z_0$. Halos were ascribed a stellar fraction as a function of their mass. However, this model is not used in \textsc{Licorice}. In the following subsection, we explain how we allow \textsc{Beorn} to use the sources of \textsc{Licorice}, thus bypassing having to use an explicit, analytical halo growth model in \textsc{Beorn}.

\subsection{Transferring \textsc{Licorice} sources to \textsc{Beorn}}\label{sec:bin_proc}

We wish to paint the 1D \textsc{Beorn} temperature profiles around the sources of \textsc{Licorice} simulations.  To do so, we process some of the outputs of \textsc{Licorice}, namely 3D cubes of UV luminosity, into acceptable inputs for \textsc{Beorn}. 
The original version of \textsc{Beorn} takes halo catalogues as inputs, sorts these halos into an arbitrary number of mass bins, and assumes a source model for each halo mass bin. We do not have the halo catalogues of \textsc{Licorice} simulations, but the \textsc{Licorice} source model does not include any dependence on the halo mass. For instance, this means that the simulation will not distinguish one halo of mass $M_h$ or ten halos of mass $M_h/10$ at the same location. Hence, we can reasonably use our 3D luminosity cubes as a source catalogue: each non-zero pixel can be perceived as a single source, even though it is likely to represent a halo population in the simulation.


\begin{figure}
    \centering
    \includegraphics[scale = 0.5]{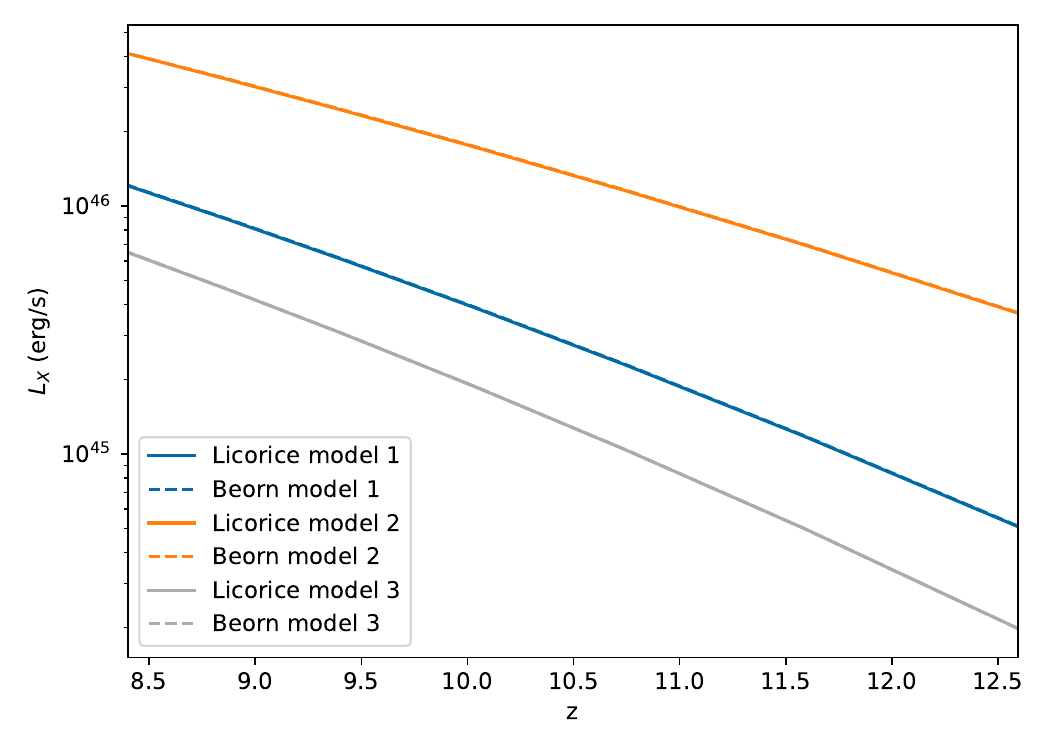}
    \caption{Total X-ray luminosity in the \textsc{Beorn} and \textsc{Licorice} versions of three different models. }
    \label{fig:lumX}
\end{figure}

We find that the pixels of the \textsc{Licorice} luminosity cubes do not follow simple analytical prescriptions such as an exponential halo mass accretion rate, as assumed in \textsc{Beorn}. Indeed, in theory, each  \textsc{Licorice} source has its own history, depending on the specific environment in which it grows. The relation between  halo mass and star formation rate or luminosity at a given redshift in a \textsc{Licorice} simulation is not bijective. To allow for \textsc{Beorn} to process the \textsc{Licorice} sources,  we follow a similar procedure as the one outlined in \cite{Ghara2018}.

In order to sort sources depending on the luminosity they will have at the last simulated redshift, we follow sources backward in time, and ascribe them their `final' luminosities as well as their current luminosities.  The procedure we follow can be described as:

\begin{itemize}
    \item We begin at the last simulated redshift, when the cube of final luminosities is equal to the cube of current luminosities. 

    \item Assuming we know the final luminosities of the sources in a cube at redshift $z$, we move on to the next cube at $z+dz$.
    
    \item For all sources (i.e., for all non-zero pixels of the luminosity cube at $z+dz$), we check if a source is present at the same location in the cube at $z$. If it is, the high redshift source inherits the final luminosity of the low redshift source. If not, we assume that the source merged or crossed a cell border. 
    
    \item If a source of higher luminosity appeared in a neighbouring cell between $z$ and $z+dz$, we assume that it is the same source that simply crossed a cell border, and we ascribe its final luminosity to the source at $z+dz$. If not, we assume a merger occurred and the source at $z+dz$ inherits the highest final luminosity of the neighbouring cells of the cube at $z$. {If no progenitor can be found, we assume the source was born at that redshift, and we assign it a luminosity of 0 at higher redshifts.  }

\end{itemize}

Thus, we build the luminosity histories of the \textsc{Licorice} sources, and keep track of the final luminosity of each source. We can therefore construct the mean luminosity histories of the sources that will end up in a given luminosity bin at the end of the simulation. In this work, we use $N=1000$. Interestingly, increasing $N$ to $\sim 10^4$ would make \textsc{Beorn} roughly as computationally expensive as \textsc{Licorice} {, at least in the relatively low-resolution regime used here, where a \textsc{Licorice} simulation is completed in a few hundred CPU-hours. At higher resolutions, Beorn would probably remain more efficient with a large number of profiles.}

The physical modelling of \textsc{Beorn} relies on star formation rates. We turn the mean luminosity histories into star formation histories using 
\begin{equation}
    \dot{M_*} = 0.00192 \,L_{UV} / c_X,
\end{equation}

\noindent where $c_X =  3.40 \times 10^{40} \rm{ \, yr \, M\odot^{-1}  \,erg\,s^{-1}} $ is the usual proportionality factor between the X-ray luminosities and star formation rates of galaxies \citep[e.g.][]{Furlanetto2006a}.
$\dot{M_*}$ is not the actual star formation rate of the \textsc{Licorice} sources (although we have checked that they are similar), but the rates that \textsc{Beorn} needs so that its sources have the same X-ray luminosity as the \textsc{Licorice} sources.
In both \textsc{Beorn} and \textsc{Licorice}, we use a soft X-ray spectrum that follows a power-law of index 1.6 within $100 \,\rm{eV}$ and $2 \,\rm{keV}$, and we turn off hard X-rays in \textsc{Licorice}.

\textsc{Beorn} can then pre-compute the temperature profiles for each bin, and paint the profiles by finding in which bin a {source} will end up at the lowest redshift, and placing around that {source} the profile corresponding to that bin at the current redshift. {In particular, the profiles at a given redshift are painted around the location of the source at the same redshift, not the position at which the source ends up at the lowest redshift. }We do not paint around the "unborn" sources identified  when constructing the luminosity histories. Instead, we only paint the profile for a bin around the sources of that bin that already exist in Licorice, and to conserve energy, the profiles are multiplied by the ratio of the number of sources that exist at that redshift divided by the number of sources that exist at the final redshift. We explain why this normalisation is required to preserve energy conservation in Appendix \ref{sec:app_track}. 

Note that the binning procedure means that we cannot ensure that all \textsc{Beorn} sources have the same luminosities as their \textsc{Licorice} counterpart at all times. However, \textsc{Beorn} sources in a given bin will have the average luminosities of the original \textsc{Licorice} sources. We believe that this {kind of binning approximation, while not  strictly mandatory, is common in 1D RT codes as a way to compute and paint a moderate number of profiles and thus reduce computation cost. For instance, \textsc{Beorn} is used with approximately a hundred different profiles in \citep{Schaeffer2023}, and \textsc{Grizzly} is used with 377 galaxy bins (i.e. 29 stellar mass and 13 age bins) in the POLAR-II setup \citep{Ma} while 3D RT codes simulate millions of different sources with individual histories)}. Nonetheless, our method is an approximation of the halo histories, as we do not compute exact merger trees.



\subsection{Differences in the radiative transfer methods of both codes}


\begin{figure*}
    \centering
    \includegraphics[scale = 0.5]{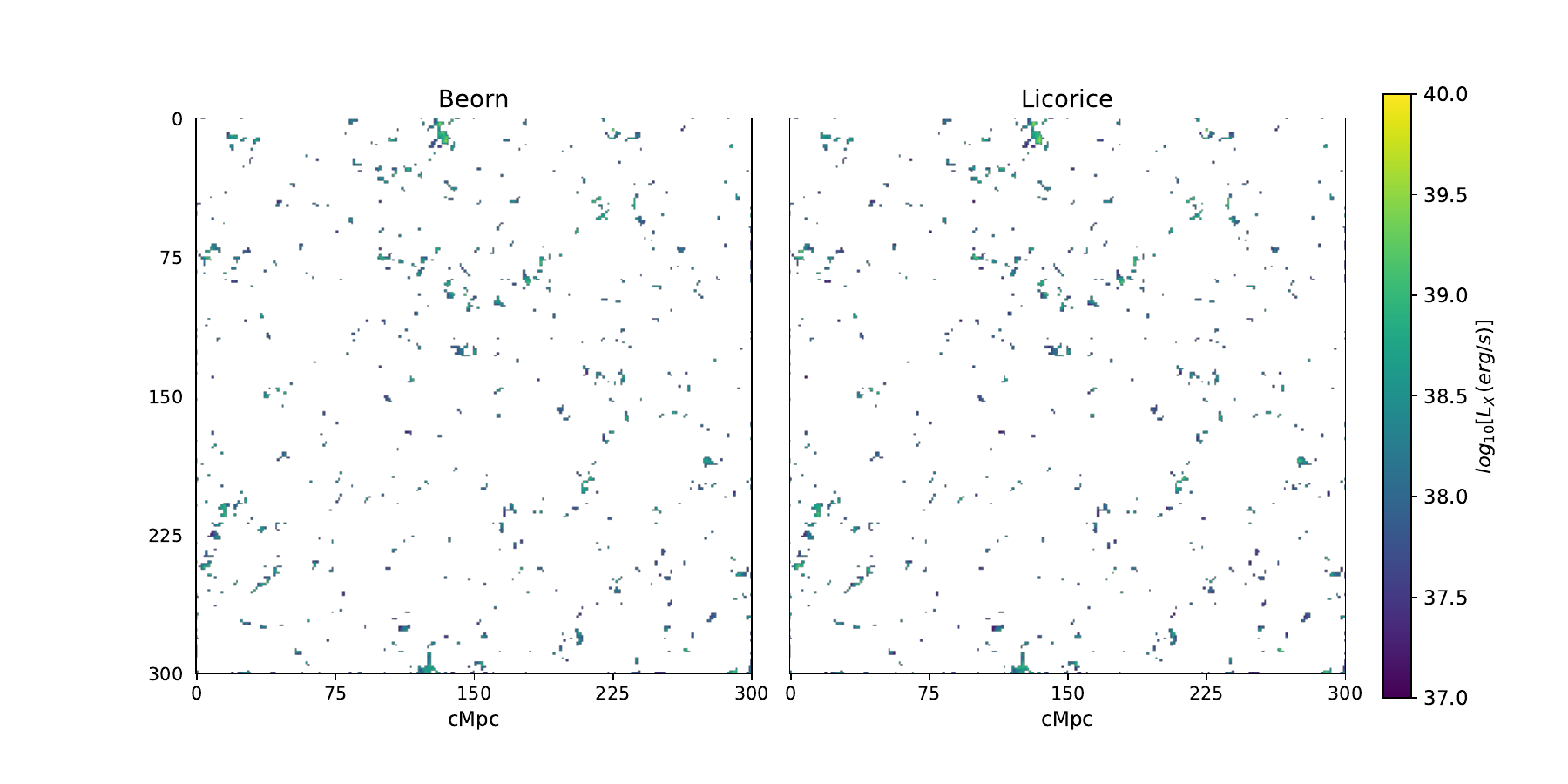}
    \caption{Slices of the luminosity field in the \textsc{Beorn} (left) and \textsc{Licorice} (right) boxes for the fiducial model at $z\sim10$, illustrating that the locations of the sources are identical in both setups.}
    \label{fig:lum_maps}
\end{figure*}

\begin{figure}
    \centering
    \includegraphics[scale = 0.55]{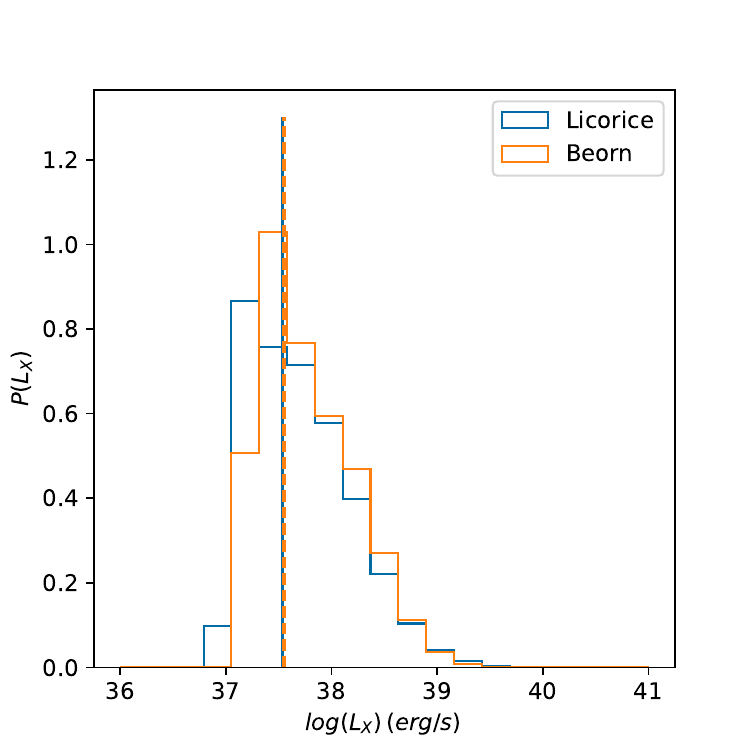}
    \caption{Luminosity distributions in the fiducial model at $z\sim10$ in the original \textsc{Licorice} box,  and the ones used by \textsc{Beorn} after the binning procedure described in Section \ref{sec:bin_proc}. {The vertical lines show the means of the distributions.} }
    \label{fig:lumX_hist}
\end{figure}

Even if the two setups use the same luminosity at all times and both sets of sources emit the same amount of X-ray photons, the result of the simulation codes will differ if the X-rays do not interact with the IGM in the same way. \textsc{Licorice} computes the number of photons absorbed by each cell crossed by a photon packet as it travels, and an absorbed X-ray deposits a fraction $f_{heat}$ of its energy as heat following equation \eqref{eq:fheat}. \textsc{Beorn} proceeds differently when the 1D temperature profiles are computed. The X-ray flux is first computed as a function of the distance from the source, as well as the fraction of X-ray absorbed by the IGM. $f_{heat}$ is then computed as a function of the ionised fraction. While \textsc{Beorn} originally used a slightly different function for $f_{heat}$, \citep{Schaeffer2023, Shull1985} we can adopt the fit of equation \ref{eq:fheat} to match the physics of \textsc{Licorice}.

However, the ionised fraction used by \textsc{Beorn} is not the local ionised fraction, since the profiles are pre-computed using only a model for the sources before being painted on a box. Instead, the code uses the {box-average} star formation rate history to estimate the free electron fraction in the neutral IGM at a given redshift, and uses this quantity to compute $f_{heat}$. 

Note that another method is used by \textsc{Grizzly}, which computes a temperature profile around a source which is assumed to be isolated{, in the sense that the profile depends only on the central source and no other}. It computes the free electron fraction in the (mostly) neutral IGM caused by the X-rays of that sole source, while \textsc{Beorn} assumes that a source exists in an `average' IGM ionised by an average collection of X-ray sources. A limit of these approximations is reached when the regions heated and ionised by different X-ray sources begin to overlap with each other, and the heating around one source leverages the ionisations caused by another. This overlap problem can significantly affect the value of $f_{heat}$ (and therefore the temperature of the IGM and the value of the 21-cm signal).

{Current 1D codes are affected by the issues caused by the overlap of X-ray fluxes.} Note that the overlap of ionised bubbles is less problematic, since cells that are `over-ionised' can be redistributed to the outskirts of the ionised region in which they lie. Still, we expect differences in the handling of this overlap to contribute to the discrepancies between the results of 1D and 3D RT codes.

Finally, we identify an additional difference between the codes: \textsc{Licorice} is an intrinsically Lagrangian code and uses the local density in the radiative transfer computations, while \textsc{Beorn} systematically uses a homogeneous density field as an approximation, except in the computation of the adiabatic temperature of the Universe. This also means that the transport of heat by moving particles in \textsc{Licorice} is absent from \textsc{Beorn}.
This approximation changes the solution of the temperature evolution differential equation, and while we can expect only a mild effect on the average temperature, it is likely to impact the morphology of the heating. X-rays seem likely to deposit their energy closer to the sources in \textsc{Licorice} compared to \textsc{Beorn}, as the vicinities of halos are likely to be overdense and overionised.

{In this section, we presented a number of modifications to the \textsc{Beorn} code that were implemented to allow a more direct comparison with \textsc{Licorice}. As discussed, these modifications mostly relate to the handling of luminous sources and slight parameter and formula adjustments to better match those of \textsc{Licorice}. The radiative transfer part of \textsc{Beorn}, which is at the heart of the comparison, was kept mostly intact. We therefore use a slightly different code to that presented in \cite{Schaeffer2023} and publicly available\footnote{https://github.com/cosmic-reionization/Beorn}, but we will continue to call that version \textsc{Beorn} throughout this work for simplicity. 

Note that while the exact implementation of the modelling choices of the radiative transfer in \textsc{Beorn} may be specific to this code, we believe that most of them belong to families of similar approximations that exist within other 21-cm simulation codes (for instance, semi-numerical codes {such as 21cmFAST or 21cmSPACE} also compute the optical depth assuming a homogeneous density field to reduce computing costs). Finally, 3D RT codes are not exact either and contain their own sets of approximations; for instance the radiative transfer from \textsc{Licorice} is affected by Monte-Carlo noise  and could suffer from the limited resolution at which we use it \citep[see e.g][]{Meriot2024}.}

\section{Results}\label{sec:results}

\begin{table*}
    \centering
    \begin{tabular}{|c|c|c|c|}
        \hline

     Case        & Mean temperature & Mean density during RT & $f_{heat}$   \\
         \hline

     \textsc{Licorice}, realistic    & mass averaged & no & local, equation \ref{eq:fheat}  \\
     \textsc{Beorn}, realistic     & volume averaged & yes & global, equation \ref{eq:fheat}   \\
     \textsc{Licorice}, idealised    & volume averaged & yes & global, 0.11 \\
     \textsc{Beorn}, idealised    & volume averaged & yes & global,  0.11  \\         \hline

    \end{tabular}
    \caption{{Key differences between the realistic and idealised cases.} }
    \label{tab:real_idealised}
\end{table*}


Below, we conduct two experiments. We use \textsc{Beorn} to post-process sources from \textsc{Licorice} simulations. In all cases, we assume that no UV photons are emitted in order to focus on the effects of X-rays. 
On the one hand, we compare the results produced by unaltered versions of the codes {(which we label `realistic')}, and on the other, we construct idealised setups to try to understand the impact of the differences identified in the previous section. {The properties of the idealised and realistic setups are summarised in Table \ref{tab:real_idealised}}.

\begin{table}
    \centering
    \begin{tabular}{|c|c|c|c|}
    \hline
      Model   &  $f_X$ & $\tau_{SF} (Gyr)$ & $log_{10}[M_{min}/M_\odot]$ \\
    \hline

       1  & 6.8  & 41 & 9.07 \\ 
       2  & 10  & 39 & 8.27 \\ 
       3  & 3  & 25 & 9.33 \\ 
       4  & 1  & 24 & 8.53 \\ 
       5  & 2  & 25 & 9.6 \\ 
             \hline

    \end{tabular}
    \caption{Astrophysical parameters of the models considered throughout this work. Model 1 is the fiducial model. {Other parameters (UV escape fraction, fraction of hard X-rays) are set to 0.}}
    \label{tab:params}
\end{table}

In the idealised setup, we make the following changes. We artificially force both simulation codes to use a constant $f_{heat}$ set to $f_{heat}(x_{ion} = 1.2 \times 10^{-4}) \approx 0.11$ at all times and places.
 We also modify \textsc{Licorice} so that the hydrogen optical depth  is computed using the average density of the box and not the local density, and solve the temperature equation on the radiative transfer grid instead of on the particles and using the average density of the simulation.  

 
 These modifications aim to mimic the homogeneous density approximation made in \textsc{Beorn}, which first paints the 1D profiles before adding $T_{adiab}(z)$ to the temperature cube.

The goal of this simplified and less realistic setup is to show the amplitude of the effects we can account for, as well as the impact of the differences that remain  unidentified or unaccounted for. For instance, the differences between the luminosity fields due to the binning procedure and the 1D aspect of the \textsc{Beorn} radiative transfer fall in the second category. 

Since we ignore UV photons and only consider soft X-rays, the relevant astrophysical parameters that we vary in the following are the X-ray emissivity $f_X$, the gas-to-stars conversion timescale $\tau_{SF}$, and the minimum mass of star-forming halos $M_{min}$. The parameters of the 5 models we consider are reported in Table \ref{tab:params}. {Those models were chosen to consider different star formation scenarios and while we use different X-ray intensities, we focus on relatively high values of $f_X$ so that the differences in the X-ray heating of both simulation codes are visible.} In the following, we use model 1 as a fiducial model.

\subsection{Luminosities in \textsc{Licorice} and \textsc{Beorn}}

We first compute the average X-ray luminosities in both the original \textsc{Licorice} simulation and its \textsc{Beorn} processed version. 
The results for the first three models listed in Table \ref{tab:params} are shown in Figure \ref{fig:lumX}. 

The difference between the total luminosities in these models is of order $1\%$, an order of magnitude consistent across all of our tests, and is likely due to the luminosity bin width used.  We are therefore confident that the total energy injected by the sources in both simulations is virtually identical. 

The sources are also placed at the same locations in both simulations, which we illustrate with a slice of the source field shown in Figure \ref{fig:lum_maps}. Again, note that although the average luminosity in both setups is the same, the binning procedure is such that we can only ensure that any single source will have exactly the same luminosity in both boxes in the limit of a large number of bins. Indeed, we show in Figure \ref{fig:lumX_hist} the distribution of luminosities in both the \textsc{Beorn} and \textsc{Licorice} luminosity cubes at $z\sim10$. The averaging procedure needed to compute a limited number of 1D RT profiles leads to a narrower distribution with fewer very bright and very dim sources. We can expect that it might cause a {difference in the distribution and scale-dependence of heat}.

\subsection{Temperatures and 21-cm signal in both setups}

\subsubsection{Average quantities}\label{sec:av_tp}


\begin{figure}
    \centering
    \includegraphics[scale = 0.45]{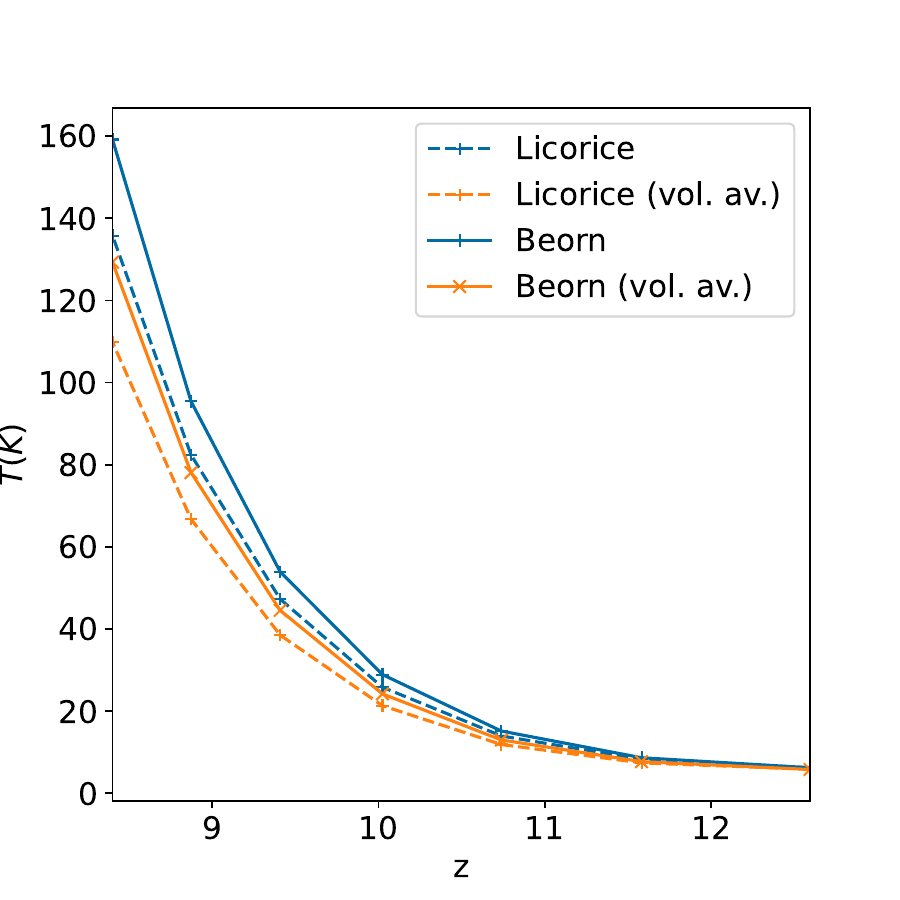}
    \caption{Mass- and volume-averaged temperatures computed in the \textsc{Licorice} and \textsc{Beorn} box for the fiducial model. }
    \label{fig:tk_vol_mass}
\end{figure}

\begin{figure*}
    \centering
    \includegraphics[scale = 0.45]{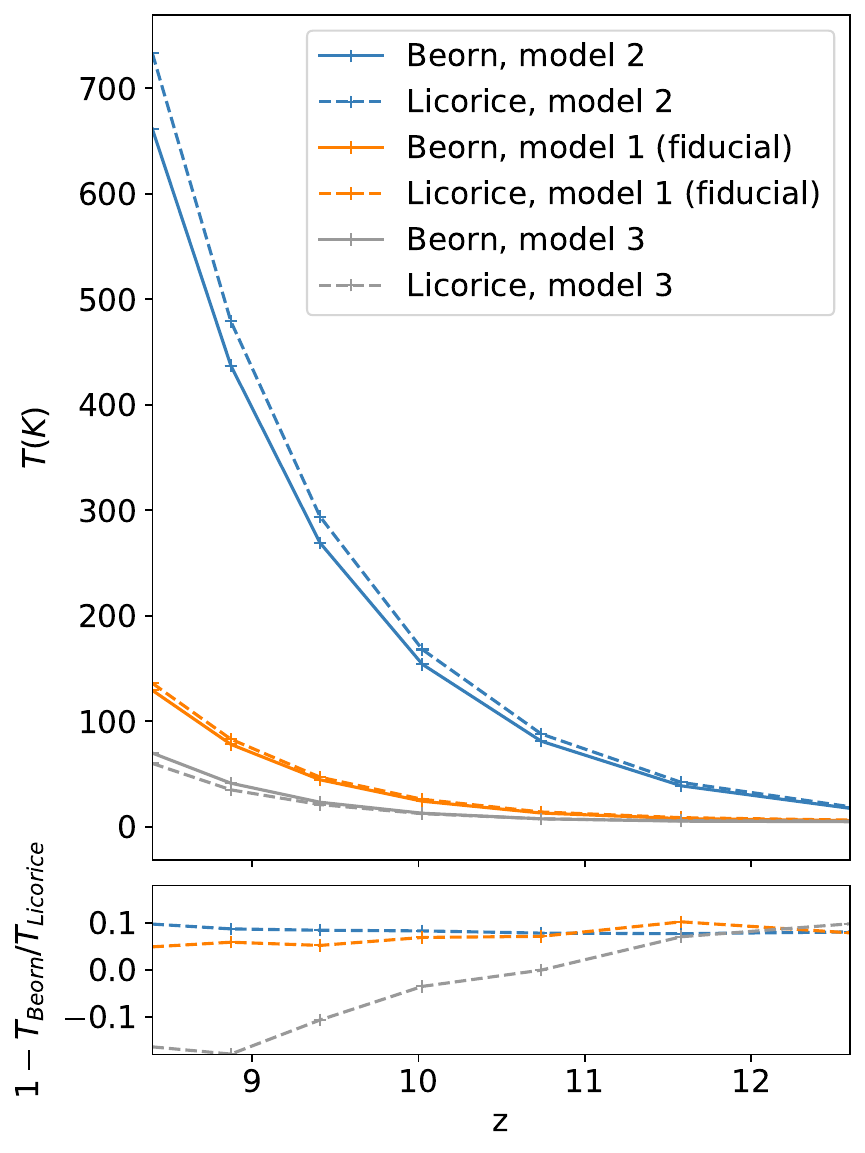}
     \includegraphics[scale = 0.45]{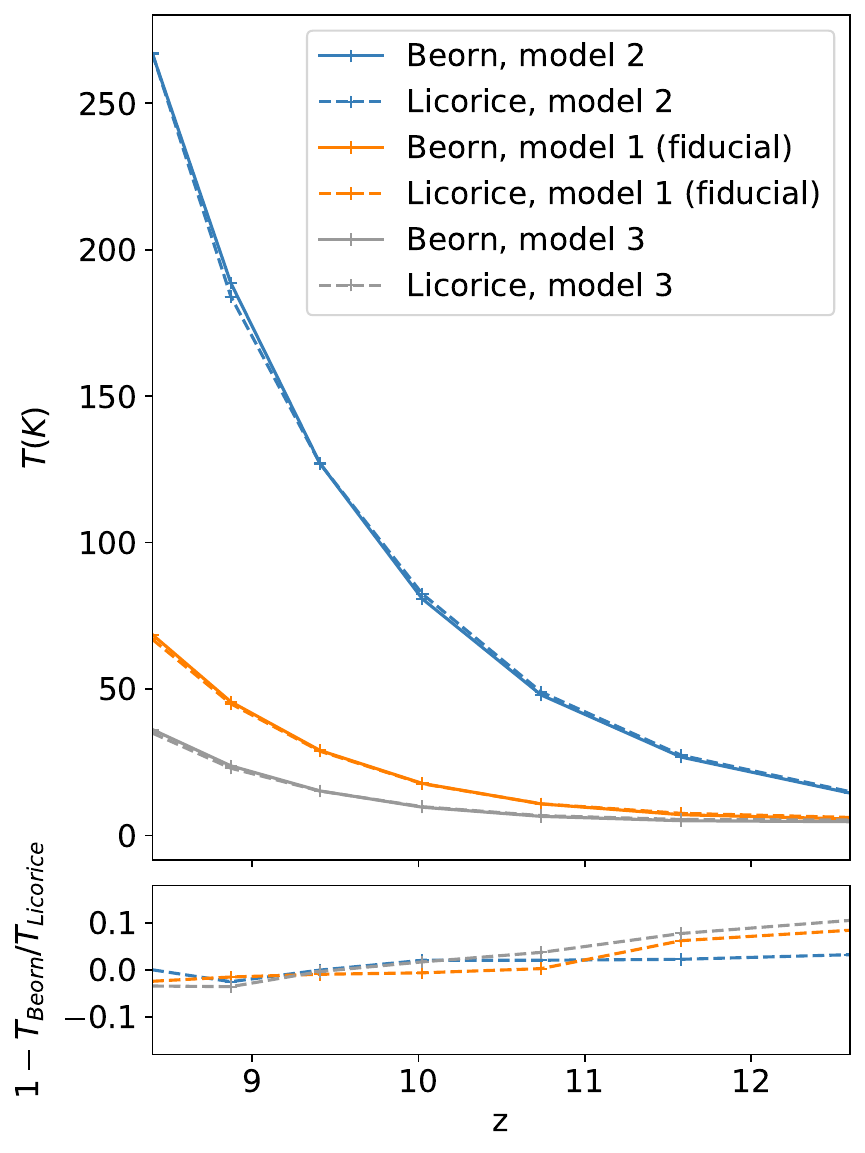}
    \caption{ \textit{Left: }Average temperature in the realistic {(i.e. without additional approximations)} case of the \textsc{Licorice} and \textsc{Beorn} versions of three models. Sources are prevented from emitting UV, so the gas is only heated by X-rays.  \textit{Right: }Average temperature in the idealised case of the \textsc{Licorice} and \textsc{Beorn} versions of three models. {The bottom panels show the relative difference between the \textsc{Licorice} and \textsc{Beorn} versions.}  }
    \label{fig:manyTk}
\end{figure*}

\begin{figure}
    \centering
    \includegraphics[scale = 0.55]{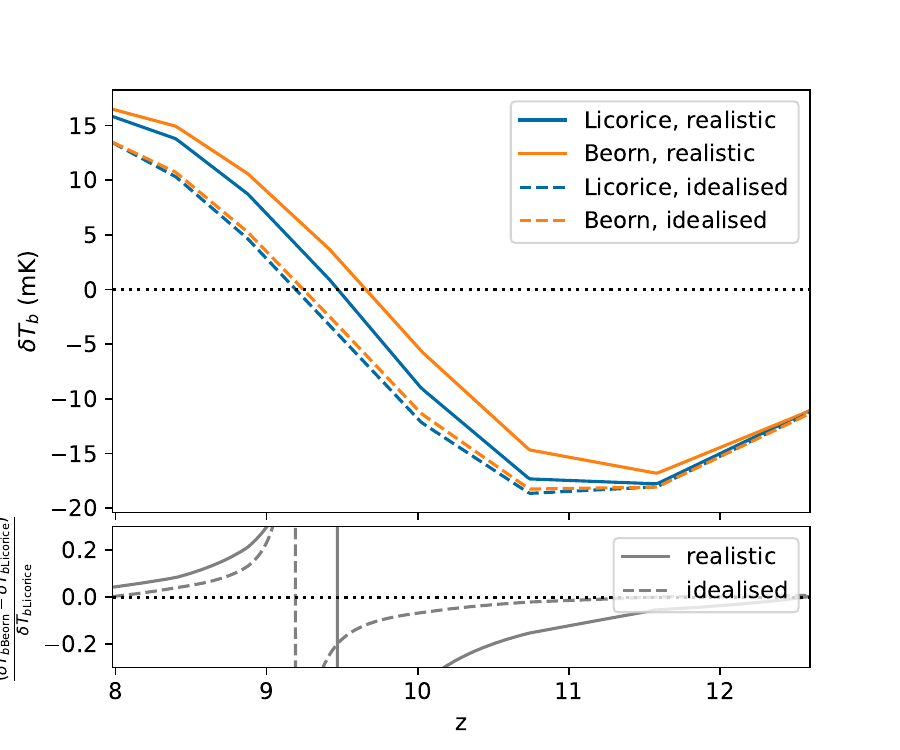}
    \caption{\textit{Top panel:} Average 21-cm signal in the realistic (solid lines) and idealised (dashed lines) setups for the \textsc{Licorice} and \textsc{Beorn} version of the fiducial model. {We also plot the relative difference for both cases in the lower panel.}  }
    \label{fig:global_signal}
\end{figure}


 {We wish to study the `Epoch of Heating', i.e. the period when the heating by X-rays is significant, but the spin temperature does not saturate yet and the Universe is assumed to be still largely neutral. Hence, we focus on redshifts between $\sim12$ and $\sim8$.}

We first want to check that the temperature of the IGM is the same in both setups. 
A difficulty lies in the computation of the average temperature itself. In \textsc{Licorice}, we can compute the volume-averaged temperature at some redshift by simply computing the average temperature of the cells of the cube output by the simulation. The mass-averaged temperature is either computed by averaging the temperature of the particles of the simulation, or by computing the density-weighted average temperature of the cells. These mass-averaging methods agree very closely, and the volume-averaged temperature {(i.e. the unweighted average of the temperature of the cells)} is typically $\sim10-20\%$ lower than the mass-averaged one.  

The situation is slightly different in \textsc{Beorn}: the temperature profiles are computed and painted assuming a homogeneous density field, an edge case where the volume- and mass-averaged temperatures are identical. Still, the density-weighted average temperature (using a real density field) is $\sim20\%$ higher than the unweighted average. To illustrate this, we show the unweighted and density-weighted temperatures of a \textsc{Licorice} simulation and its \textsc{Beorn} version in Figure \ref{fig:tk_vol_mass}. {In both cases, the volume-averaged temperatures are $10-20 \%$ lower than the mass-averaged temperatures.}

Since we are trying to check that the sources inject the same amount of energy into the IGM in both codes, {we first look at the idealised setups and} we compare in the right panel of Figure \ref{fig:manyTk} the {unweighted} temperature of \textsc{Licorice} with the unweighted temperature of \textsc{Beorn}.

We observe a {very good} agreement between the temperatures computed by \textsc{Beorn} and \textsc{Licorice}, {with differences of order $\lesssim  5\%$. Since the source populations also reach a generally good agreement, this is a validation of the pipeline we use to run \textsc{Beorn} over \textsc{Licorice} sources, and shows that energy  is conserved in both codes. 

Note that we initially processed the \textsc{Licorice} sources differently {(by allowing the number of sources in a given luminosity bin to vary over time)}, which led to \textsc{Beorn} violating energy conservation at the profile painting stage and caused a $10-30\%$ difference between the mean temperatures, depending on the underlying model.  We discuss this in more detail in Appendix \ref{sec:app_track} {and Appendix \ref{app:dep_power}.} 



Of course, a constant $f_{heat}$ is unrealistic. {We therefore move on to the realistic setup, where we compare the codes without the aforementioned approximations: \textsc{Licorice} uses its inhomogeneous density field and both codes use  the $f_{heat}$ of equation \eqref{eq:fheat}. } We show in Figure \ref{fig:manyTk} the temperatures of the same simulations, {and this time compare the density weighted average temperature of \textsc{Licorice} with the unweighted temperature of Beorn {(as those are the quantities that represent the internal energy of the IGM in the corresponding simulation codes)}. } The \textsc{Beorn} code was given the average electron fraction $\langle x_e \rangle$ computed in \textsc{Licorice} {in order to compute $f_{heat}$}. The agreement of the IGM temperature in this case is {slightly worse than} the one observed in Figure \ref{fig:manyTk}, {with differences of $\sim10 \%$}.



{The observable probe of the EoR that \textsc{Beorn} and \textsc{Licorice} ultimately simulate is the 21-cm signal of neutral hydrogen.} {We naturally compute it using the canonical equation \citep[e.g.][]{Furlanetto2006a}:}
\begin{equation}\label{eq:dtb}
\begin{split}
        \delta T_b & = 27. \, x_{HI}(1+\delta) \left[ \frac{T_s - T_{CMB}}{T_s} \right] \left[ 1 + \frac{dv_{||}/dr_{||}}{ H(z) }  \right]^{-1}  \\ 
    & \times \left[ \frac{1+z}{10}\right]^{1/2} 
    \left[ \frac{\Omega_b}{0.044} \frac{h}{0.7} \right] 
    \left[ \frac{\Omega_m}{0.27}\right]^{1/2} \, \rm{mK},
 \end{split}
\end{equation}
{\noindent where $x_{HI}$ is the neutral fraction (here set to 1), $T_S$ the spin temperature, $T_{CMB}$ the CMB temperature, $dv_{||}/dr_{||}$ the velocity gradient.}

We show the average 21-cm brightness temperature in both boxes in Figure \ref{fig:global_signal}. In doing so, we use the same Lyman-$\alpha$ coupling computed {on the \textsc{Licorice} sources} using the 1D code \textsc{SPINTER} \citep{Semelin2023a}. We observe a {good agreement} between the two global signals, {although the difference is} apparently larger than between the average temperatures. This suggests that the spatial fluctuations of the heating may be different in both setups: different correlations between the temperature field and the Wouthuysen-Field coupling or the density field would result in a different average brightness temperature, {and a study about the dependence of global signal on the correlations between these different fields can be found in \cite{Schaeffer}}. 
{Note also that the computation of the 21-cm signal applies to a volume element, and Figure \ref{fig:tk_vol_mass} shows that in the realistic case the volume-averaged temperature in \textsc{Beorn} agrees with the mass-averaged temperature of \textsc{Licorice} and not with its volume-averaged temperature, as heating is density-dependent in \textsc{Licorice}.} Still, we obtain an excellent agreement in the idealised case, {in which the codes use identical approximations}.

\subsubsection{Field evolution}

\begin{figure}
    \centering
    \includegraphics[scale = 0.55]{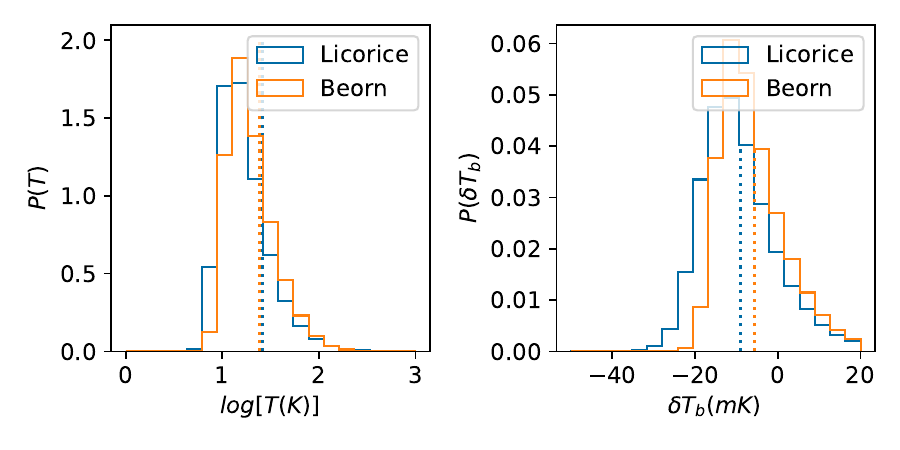}
    \includegraphics[scale = 0.55]{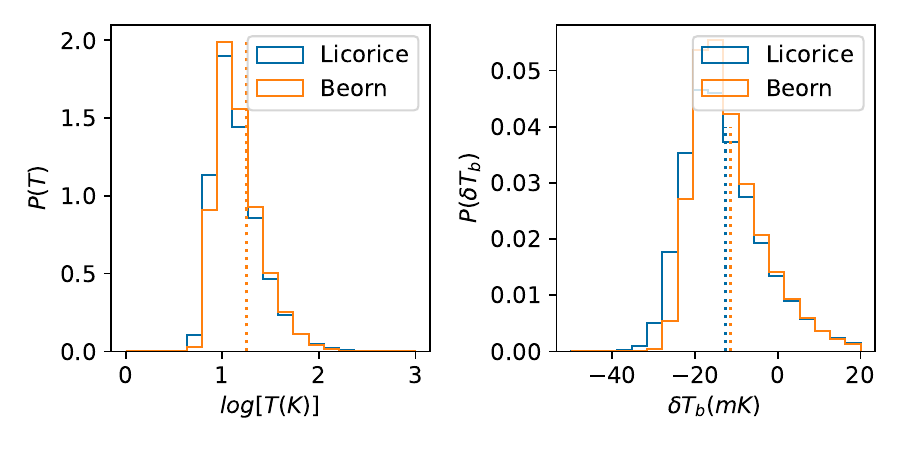}
    \caption{Distribution of temperature and 21-cm signal in the \textsc{Licorice} and \textsc{Beorn} fiducial model at $z\approx10$. \textit{Top:} realistic setup. \textit{Bottom:} idealised setup.}
    \label{fig:temp_pdf}
\end{figure}

\begin{figure}
    \centering
    \includegraphics[scale = 0.45]{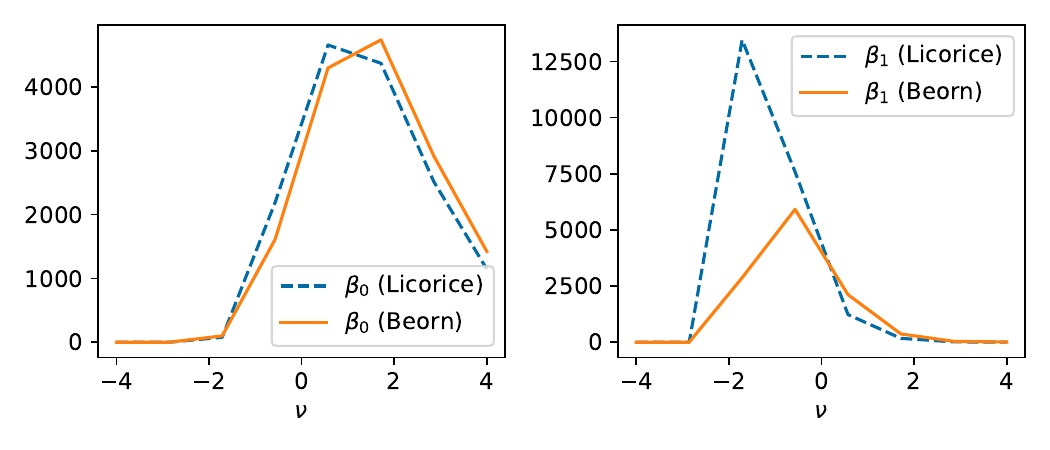}
    \caption{First two Betti numbers of the \textsc{Licorice} and \textsc{Beorn} fiducial model at $z\approx10$ in the realistic case.  }
    \label{fig:betti}
\end{figure}

\begin{figure}
    \centering
    \includegraphics[scale = 0.55]{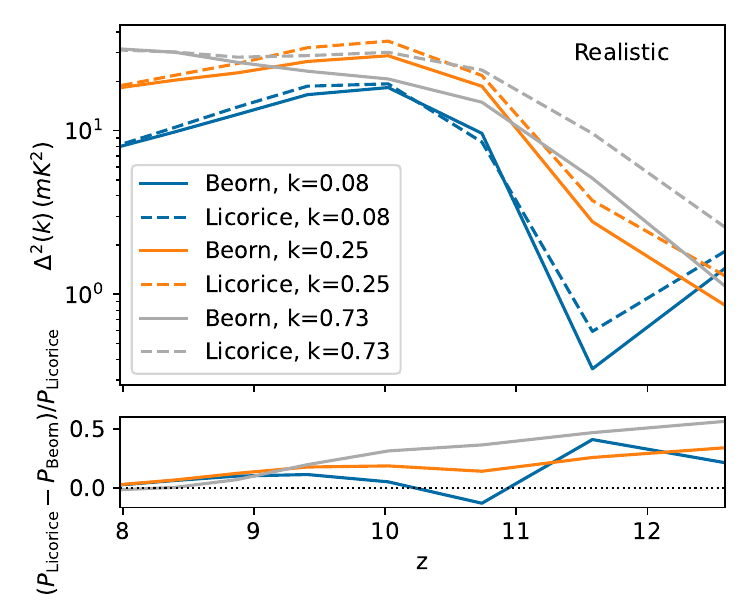}
    \includegraphics[scale = 0.55]{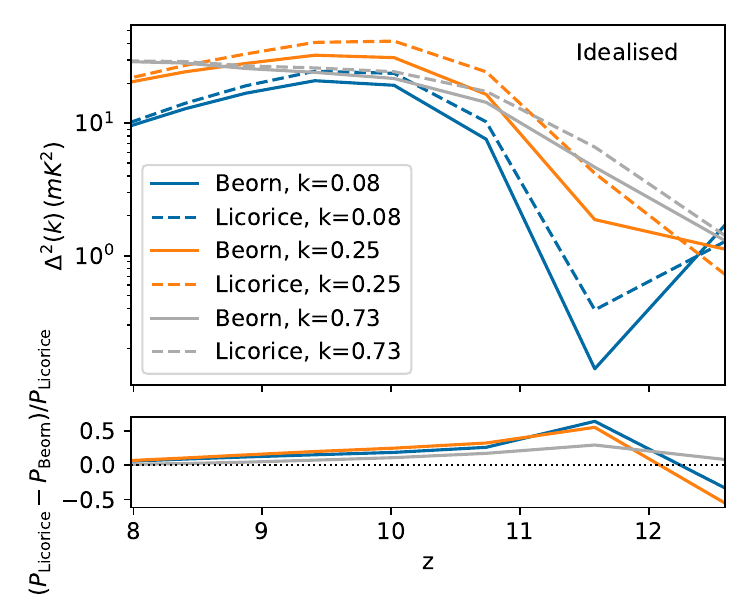}
    \includegraphics[scale = 0.5]{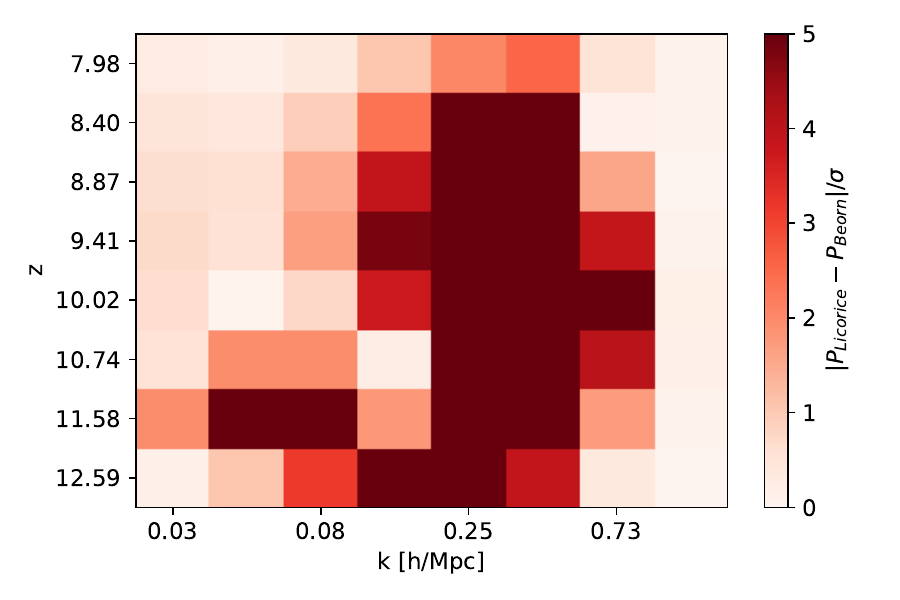}
    \caption{\textit{Top:} { {Power spectra as a function of z of the realistic fiducial model simulated by \textsc{Beorn} and \textsc{Licorice}. The spectra are plotted at three values of the mode $k$ (expressed in units of $h/\Mpc$). The sub-panel shows the relative difference}. \textit{Middle:} same, but for the idealised fiducial model. \textit{Bottom: } the absolute value of the difference between \textsc{Beorn} and \textsc{Licorice} compared to $\sigma$, the total standard deviation of the noise (i.e. cosmic variance and thermal noise corresponding to 100h of SKA observation).  {The colour bar was truncated at $5\sigma$} for visualisation. }}
    \label{fig:powerspectrum}
\end{figure}

\begin{figure}
    \centering

    \includegraphics[scale = 0.5]{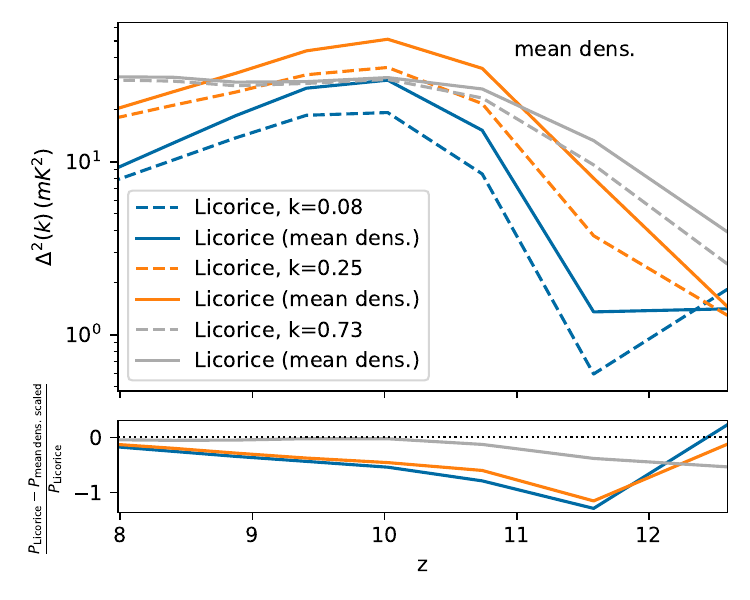}
    \includegraphics[scale = 0.5]{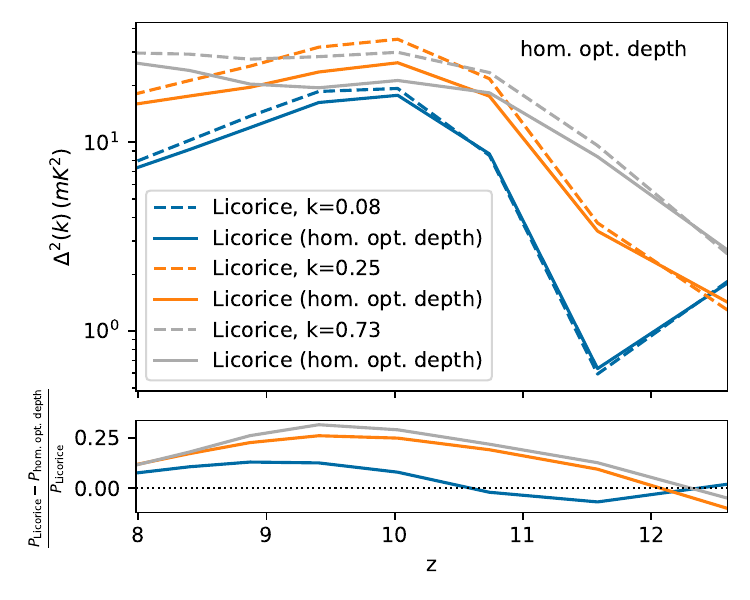}
        \includegraphics[scale = 0.5]{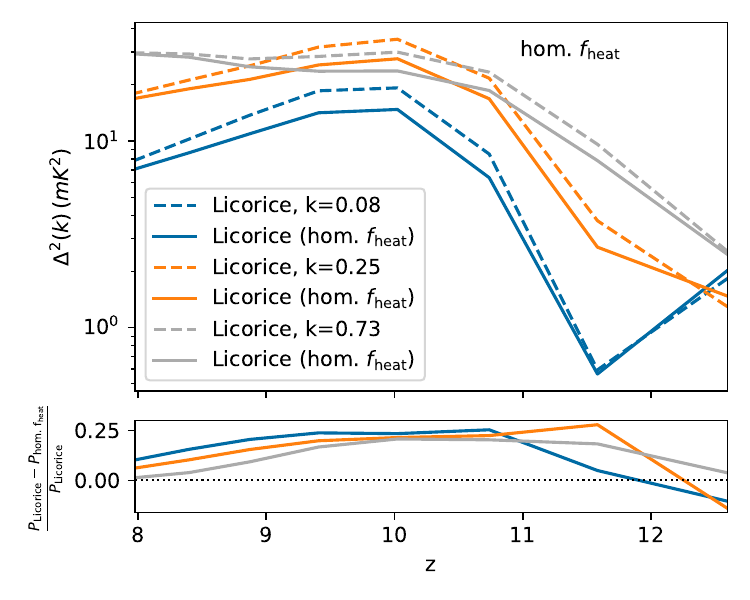}
         \includegraphics[scale = 0.51]{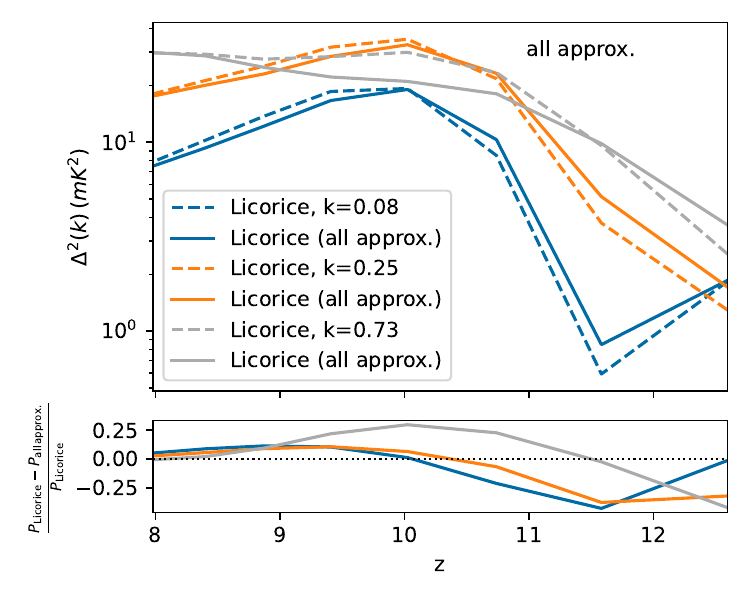}
    \caption{ { Comparison between the power spectrum computed in the realistic fiducial model (solid) and  computed with the same astrophysical parameters but assuming additional physical approximations (dashed)}. \textit{Top:} mean density is assumed in the optical depth calculation and the temperature evolution equation. \textit{Second :} we make the mean density optical depth computation but the temperature computation still relies on the true density field. \textit{Third :} we use the globally averaged value of $x_{ion}$ when computing $f_{heat}$ during radiative transfer.  {\textit{Bottom : } We assume both mean density during RT and homogeneous $f_{heat}$. } }
    \label{fig:pkratio_meadens}
\end{figure}

We now move from aggregate quantities and consider the coeval cubes produced by both codes.
We show slices of the temperature cubes, computed with no artificial approximations in \textsc{Licorice} (i.e. with a variable $f_{heat}$) in Appendix \ref{app:slices}.

{Qualitatively, the temperature fields look very similar, as sources of close luminosities are located at the same positions in both setups and the average temperatures are close. Generally, \textsc{Licorice} seems to display slightly more small scale fluctuations. To characterise the fields, we also plot the probability density distributions of the temperature and 21-cm signal in Figure \ref{fig:temp_pdf} at $z\approx10$. The distributions have similar means, as discussed before but are not identical. In Beorn, the temperature distributions are narrower, {show fewer} very hot and very cold pixels and {a larger number of} intermediate values, hinting at a different heat distribution in the box. 

Finally, following \cite{Cerardi2025}, Figure \ref{fig:betti} shows the first two Betti numbers of the \textsc{Licorice} and \textsc{Beorn} realistic cases of the fiducial model as a function of a threshold defined by $\nu = \delta T_b / \sigma_{T_b}$, where $\sigma_{T_b}$ is the standard deviation of the brightness temperature cube. The first two Betti numbers measure topological properties of the field: $\beta_0$ counts the number of isolated components above the threshold, and  $\beta_1$ the number of loops (or holes). While there is high level of agreement between the $\beta_0$, $\beta_1$ is consistently higher in Licorice, showing again that the distribution of heating is significantly different between the two codes.
We believe the differences shown here could be caused by the processing necessary for \textsc{Beorn} to process the \textsc{Licorice} sources and paint a limited number of profiles. Overall, the distributions show a generally good agreement in the realistic case, given the structural differences of the two codes.}

\subsubsection{Power spectra}

To further quantify the differences seen until now, we compute the power spectra of the 21-cm cubes in both the constant $f_{heat}$ and in the true $f_{heat}$ setups. {The {power spectra and } relative difference between the \textsc{Beorn} and \textsc{Licorice} versions are shown in Figure \ref{fig:powerspectrum}, in both the realistic and idealised setups, as well as the ratio of the difference and the standard deviation of the noise of the setup $\sigma = \sqrt{\sigma_{thermal}^2 + \sigma_{\rm{CV}}^2 + \sigma_{\rm{train}}^2}$ (see Section \ref{sec:inference} for details).} 

{ The mean absolute difference is $20 \%$  in the realistic setup and  $11 \%$  in the idealised setup.  } {We show the temperature power spectra in Appendix \ref{sec:app_t_pk}; they also agree at a $\sim20\%$ level but show different evolutions with redshift and scale. In both the realistic and idealised cases, the agreement between the 21-cm power spectra improves at lower redshift, as the IGM gets strongly heated and the brightness temperature saturates. The differences are stronger at higher redshift, around $z\sim11-12$, near the absorption peak of the global signal. At these redshifts, the heating becomes a driver of the 21-cm evolution and the temperature fluctuations are strongly correlated to the Lyman-$\alpha$ coupling and the density field. Those ingredients are coupled in non-trivial ways, and the modifications made in the idealised setup fail to fully erase the differences between the Beorn and Licorice power spectra, even though the agreement between the mean temperatures and global signals is better. We observe differences on scales larger than the typical Beorn temperature profile, which suggests an impact of the preprocessing of the sources. }

The differences observed in the realistic setup are significant, as they exceed the amplitude of cosmic variance on large scales and the instrumental thermal noise on moderate scales. Differences such as the one we show here, are therefore certain to noticeably impact inference results, {as they translate to an average difference of $4 \sigma$ for the fiducial model}.

To try to root out the causes of the observed discrepancy, we compare the power spectrum of the fiducial realistic model and power spectra computed in \textsc{Licorice} simulations with the same parameters but run with additional approximations. {We consider four cases : }

\begin{itemize}
    \item {We apply the homogeneous density field approximation (but, as opposed to the idealised setup, $f_{heat}$ is computed using Equation \ref{eq:fheat})}
    \item {We use the mean density approximation in the optical depth computation only, not in the temperature evolution equation.}
    \item {We compute the X-ray heating using the globally averaged value {of the free electron fraction}. That is, we use $f_{heat}(\langle x_e \rangle)$ in Equation \ref{eq:heat_rate_lico}, and unlike the idealised setup, it is homogeneous but not constant. }
    \item {We run a simulation with both the mean density field approximation and a homogeneous $f_{heat}$.}
\end{itemize}

{We plot the power spectra and relative errors in Figure }\ref{fig:pkratio_meadens}.
In all cases, we obtain differences of a few dozen percent, although they do not necessarily {map in a direct way} to those plotted in Figure \ref{fig:powerspectrum}. This seems especially true {around redshift $11-12$}. {The setup with all approximations reproduces the realistic \textsc{Beorn} results most closely except at redshift $\sim12$ }and we interpret this difference as the non-trivial mix of the different approximations tested, coupled to the impact of the binning procedure required for \textsc{Beorn} to compute its profiles and the absence of heat transports by the \textsc{Licorice} particles in \textsc{Beorn}. {However, this figure suggests that the individual physical approximations  are likely to cause significant differences on the 21-cm power spectra (and subsequent inference results) on their own}.

\begin{figure*}
    \centering
    \includegraphics[scale = 0.7]{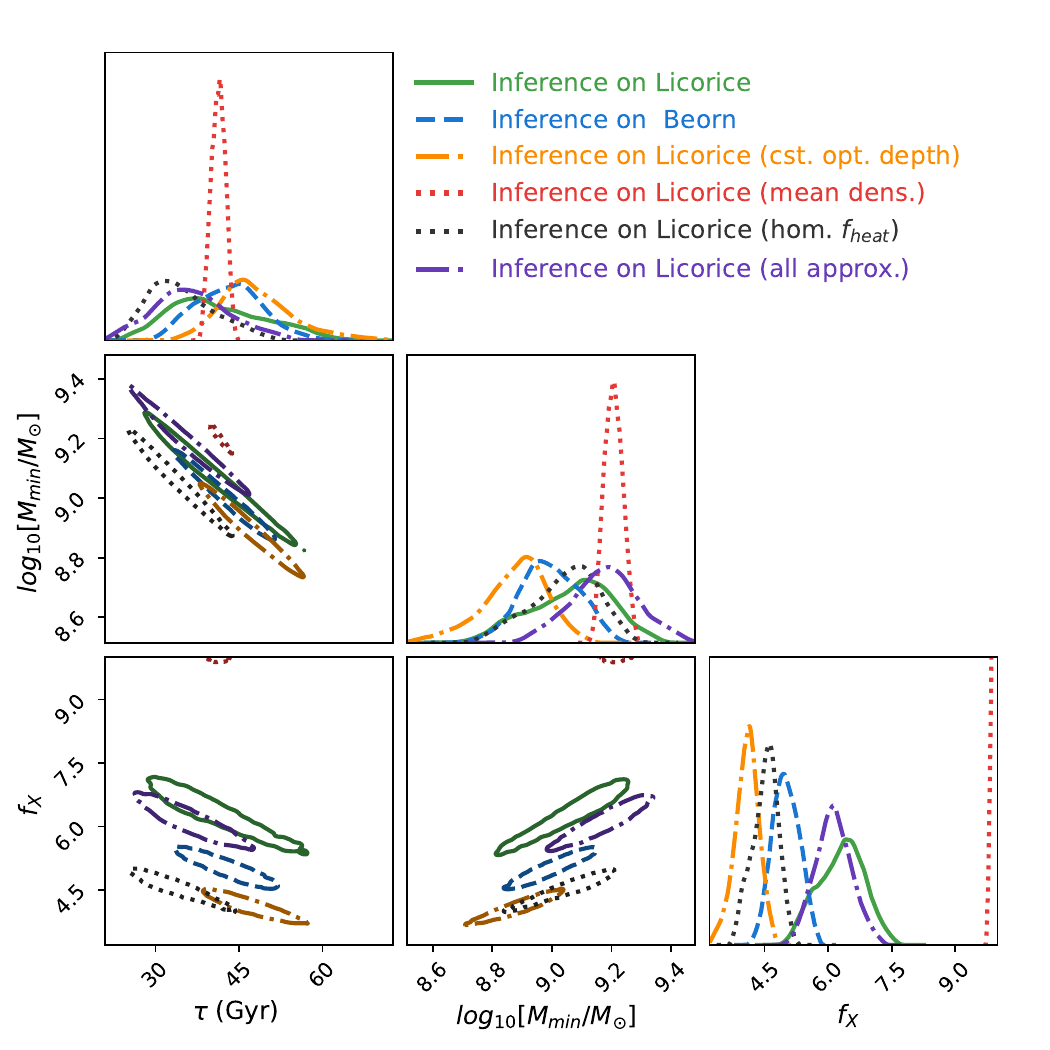}
    \caption{Posteriors obtained after performing inferences with Loremu II on the \textsc{Licorice} power spectrum of the fiducial model (\textit{green, solid}) and the corresponding \textsc{Beorn} power spectrum (\textit{blue, dashed}). Inferences on \textsc{Licorice} with the mean density approximation (\textit{red, dotted}), a homogeneous optical depth (\textit{orange, dash-dotted}), and a homogeneous $f_{heat}$ (\textit{black, dotted}) are also displayed.
    Only the $2\sigma$ contours are displayed on the 2D panels. The inferences were done with a noise level that corresponds to 100 hours of observations with SKA. }
    \label{fig:inference_both}
\end{figure*}

\subsection{Impact on inference}\label{sec:inference}


To provide an estimate of the impact of the differences we observed on our ability to recover astrophysical parameters, we re-use Loremu II, the emulator of \textsc{Licorice} power spectra presented in \cite{Meriot2025} to perform inference first on the power spectrum of the fiducial \textsc{Licorice} model, then on its \textsc{Beorn} counterpart. {We also run the inference on the power spectrum predicted by \textsc{Licorice} in the four cases described previously : when the homogeneous density field is used (either during all of RT or in the optical depth only), when the global value for $f_{heat}$ is used, and finally on the version that includes both approximations}. The former will provide posteriors that will serve as our ground truth, and the others aim to incorporate the modelling differences between \textsc{Licorice} and \textsc{Beorn} and to evaluate the impact of individual approximations. We summarise the approximations made in each case in Table \ref{tab:approx_inf}.  

\begin{table}
    \centering
    \begin{tabular}{|c|c|c|c|}
        \hline

     Case        & Optical depth & Mean density & $f_{heat}$   \\
         \hline

     \textsc{Licorice}    & local & no & local  \\
     \textsc{Beorn}       & global* & yes* & global*  \\
     \textsc{Licorice} (mean dens.)   & global* & yes* & local \\
     \textsc{Licorice} (hom. opt. depth)   & global* & no & local  \\
     \textsc{Licorice} (hom. $f_{heat}$)  & local & no & global* \\

          \textsc{Licorice} (all approx. )  & global* & yes* & global* \\
         \hline

    \end{tabular}
    \caption{Approximations made in all cases considered throughout Section \ref{sec:inference}. * indicates a physical approximation. }
    \label{tab:approx_inf}
\end{table}

To obtain the posteriors, we use the \textsc{Licorice} emulator in an MCMC pipeline. We use an explicit Gaussian likelihood $L$ defined by 
\begin{equation}\label{eq:diag_likelihood}
    logL(y | \theta) = - \sum_{k,z} \frac{1}{2} \left( \frac{P_{21}(k,z) - y(\theta, k ,z )}{\sigma_{\rm{tot}}(k,z)} \right)^2,
\end{equation}

\noindent where $\theta$ is the parameter vector, $P_{21}$ the simulated power spectrum that is used as an inference target, $y$ the output of the emulator. $\sigma_{\rm{tot}}$ represents the variance of the problem, and is defined by 
\begin{equation}
        \sigma_{\rm{tot}} = \sqrt{\sigma_{thermal}^2 + \sigma_{\rm{CV}}^2 + \sigma_{\rm{train}}^2}.
    \end{equation}
    
In this expression, $\sigma_{thermal}$ is the standard deviation of the instrumental thermal noise. We consider a noise level corresponding to 100 hours of SKA observation. {Similarly to \cite{Meriot2025, Semelin2024a} we assume an instrument with the 512 35m stations of the SKA 2016\footnote{\url{https://www.skao.int/sites/default/files/documents/d18-SKA-TEL-SKO-0000422_02_SKA1_LowConfigurationCoordinates-1.pdf}} base design, with an effective collective area of $min(2.56, \lambda^2) \, \rm{m}^2$, system temperature $T_{sys} = 100 + 300\left( \frac{\nu}{150 \rm{MHz}} \right)^{-2.55}$ K, with perfect foreground removal. The design of the instrument has slightly evolved since, but this should not affect the qualitative conclusions of this work.} $\sigma_{\rm{CV}}$ corresponds to the Cosmic Variance of our simulation, and is, as $\sigma_{thermal}$, computed following \cite{Mcquinn}. $\sigma_{\rm{train}}$ represents the emulator error, and we refer to \cite{Meriot2024} for more details regarding its calculation. We run the inference on our fiducial model, with parameters mentioned in Table \ref{tab:params}. {The other parameters (UV escape fraction and hard X-ray fraction) are set to their true value, 0, during the inference.}

{Note that in this section, we vary the targets of inference, whereas ideally we would run inferences on the same target with different emulators (one for each physical approximation and for the \textsc{Beorn} version of the \textsc{Licorice} boxes). Naturally, this requires as many simulated training sets as we want emulators, which is computationally expensive and beyond the scope of this work. Still, if the differences between the different versions are small compared to the sensitivity of the instrument, we expect the posteriors to be similar. }    

The posteriors are shown in Figure \ref{fig:inference_both}. We observe a significant difference between the 1D posteriors obtained on the \textsc{Licorice} signal and the ones that include approximations. Including these approximations offsets the 1D posterior of all parameters by typically $\gtrsim 0.5-1 \sigma$, and the effect is the strongest on $f_X$, for which discrepancies of several $\sigma$ exist between the 1D posteriors of the studied cases. Incidentally, the "mean density" posterior is pushed to the edge of the prior on $f_X$ which probably explains its narrowness. {The 1D marginalised "all approximations" posteriors overlap with the Licorice ones, as the approximations can have opposite effects on the power spectrum. However, the 2$\sigma$ contours are practically disjoint on the 2D panels, and the apparent overlap does not hold everywhere in parameter space, as we discuss later. }

We compare the \textsc{Licorice} and \textsc{Beorn} spectra on the 5 parameter sets considered in the previous section. For each pair of 1D posteriors, we compute a relative bias on a parameter $\theta$ as follows:
\begin{equation}
    \Delta \theta = \frac{ \langle \theta_{{Licorice}} \rangle -  \langle \theta_{Beorn} \rangle }{ max(\sigma_{{Licorice}}, \sigma_{Beorn} ) },
\end{equation} 

\noindent where  $\langle \theta \rangle$ is the mean value of the 1D marginalised posterior on $\theta$ obtained after running the inference, and $\sigma $ the standard deviation of the 1D posterior obtained by the inference. {We show the relative biases in Figure \ref{fig:inf_biases}. On all 5 models, we find significant biases on nearly all parameters, ranging from $\gtrsim 0.1 \sigma$ to several $\sigma$, both when comparing the Licorice posteriors to the Beorn or "all approximations" Licorice ones. } {The differences in terms of absolute values of the parameters, however, are mostly contained to $\lesssim 10 \rm{Gyr}$, $\lesssim 0.2$, and $\lesssim 1$ for $\tau_{SF}$, $log_{10} M_{min}$, and $f_X$ respectively, as the extraordinary sensitivity of the SKA would allow us to distinguish between models that are physically reasonably similar.  }

As discussed previously, these results characterise a small number of simulations and the distribution of the sources processed by \textsc{Beorn} is not identical to the one generated by \textsc{Licorice}. The results rely on a comparison between two specific codes and the source of the difference {between the power spectra} has not been entirely identified, making the biases observed hard to generalise. As such, we believe they are best interpreted as an order of magnitude. 

{Although we varied the targets and not the emulator, those results suggest} that interpreting the same data with different 21-cm codes can {currently} lead to significant discrepancies. {However, it is possible that additional parameters could be implemented in \textsc{Beorn} to model the differences with \textsc{Licorice}, and marginalising over those parameters at the inference stage may lead to a better agreement between 1D and 3D codes while still maintaining the computational cost advantage of 1D RT {(or some implementation of a modelling error, see e.g. \cite{Greig2015})}.    }

\begin{figure}
    \centering
    \includegraphics[scale = 0.45]{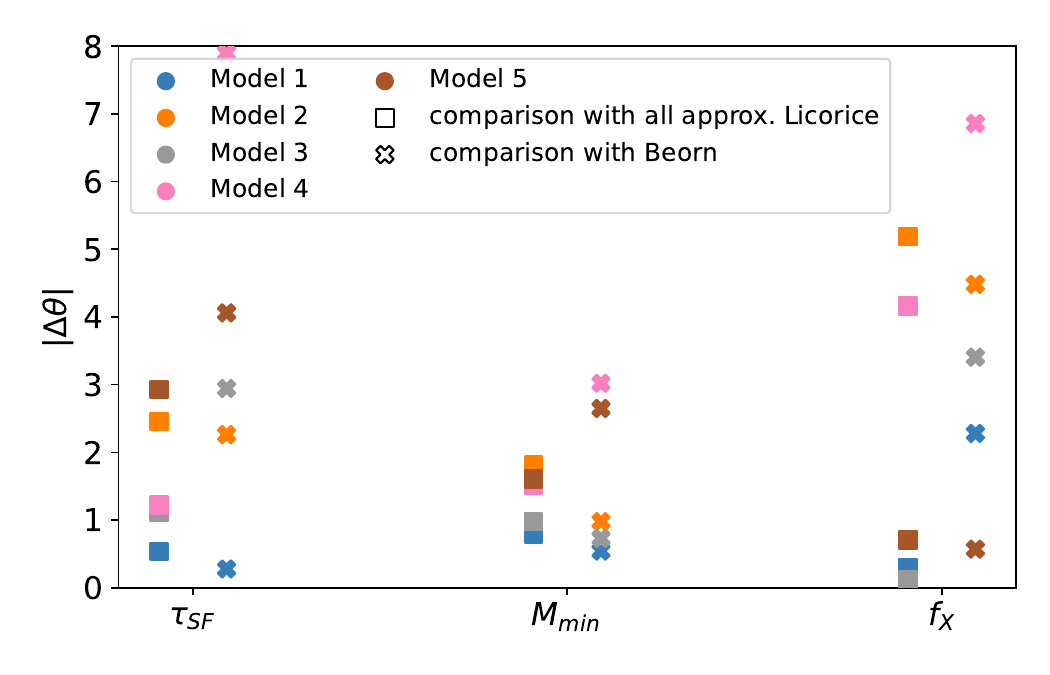}

    \caption{Relative biases {(quoted as a number of $\sigma$)} on the astrophysical parameters of the 5 considered models {when comparing the realistic \textsc{Licorice} and \textsc{Beorn} results} (\textit{crosses}) and {when comparing the realistic \textsc{Licorice} and the "all-approximations" \textsc{Licorice} results} (\textit{squares}).}
    \label{fig:inf_biases}
\end{figure}

\section{Conclusions and Discussion}\label{sec:concl}
    In this paper, we process sources simulated by the 3D RT \textsc{Licorice} with the 1D RT code \textsc{Beorn} and compare the IGM properties simulated by both codes, with the same parameters and physics. To do so, we first generate a \textsc{Licorice} simulation. We extract the luminosity cubes at each snapshot. We construct luminosity histories of the \textsc{Licorice} sources by tracking each non-zero pixel of the final luminosity cube backward in time. These histories are then sorted in bins and averaged based on their value at the final redshift. \textsc{Beorn} then uses these binned histories to compute as many temperature profiles, which are then painted around the corresponding \textsc{Licorice} sources. {While this method cannot guarantee that all the sources processed by \textsc{Beorn} have the same luminosities as they had in the \textsc{Licorice} simulation (except when using as many luminosity bins as there were sources in \textsc{Licorice}, which is very expensive in practice), we observe an excellent agreement of the total luminosity in each version of the box. The luminosity distributions agree broadly but not perfectly.}

    {We then obtain the cubes of IGM temperature computed by \textsc{Licorice}, and \textsc{Beorn} based on the same sources using the prescriptions described earlier}. We find that both sets of temperature cubes are significantly different. The average temperature follows a similar evolution, although a difference of approximately $\Delta z \sim 0.2$ can be observed. The fields themselves show differences, which we try to explain by running another set of simulations with approximations and simplifications. These include assuming a homogeneous density field in the radiative transfer module of \textsc{Licorice}, and assuming a constant $f_{heat}$ in both codes. These physical approximations {do improve the agreement between the mean temperatures of the simulations but still lead to visual differences in the fields, indicating that some fraction of the differences in the temperature fluctuations are  either due to  the binning procedure {used to process \textsc{Licorice} sources with the 1D code} or to unidentified effects.   }

        We use the \textsc{Licorice} and \textsc{Beorn} temperature cubes to compute cubes of the 21-cm signal, using the density and Wouthuysen-Field coupling fields computed by \textsc{Licorice} and \textsc{SPINTER}. Naturally, the differences between the temperature cubes are transferred to the brightness temperatures. We compute the power spectra for 5 sets of parameters, and find differences of a factor $\sim 10-50\%$ between the 21-cm power spectra. Using an MCMC pipeline relying on Loremu II, an emulator of the \textsc{Licorice} power spectra of the Loreli II simulations, we estimate that such differences in power spectra typically cause biases of $\gtrsim 1\sigma$ on the resulting posteriors. Worryingly, these model-error biases exceed the biases typically found when using different inference methods but a single simulation code. {Thanks to the great sensitivity of the SKA, however, these differences can amount to several $\sigma$ while translating to reasonably limited  parameter differences. } {Although we varied the inference targets and kept the emulator fixed in the MCMC setup, this does imply that the studied approximations are significant enough to impact the interpretation of future 21-cm measurements.}

    In addition, we run \textsc{Licorice} simulations with the fiducial parameters {but assuming either 
    a homogeneous density field during the radiative transfer calculations, an optical depth computed using the mean density of the box instead of the local density, or a homogeneous $f_{heat}$.} We compare the outputs with the realistic \textsc{Licorice} simulation and observe differences of a similar magnitude but different structure than when comparing with the \textsc{Beorn}-processed box. We perform the inferences, and again observe significant biases. This suggests that these physical approximations alone could individually be responsible for large discrepancies, aggravated by the binning procedure of the luminous sources. Note that using the mean value of the density of the Universe in the computation of the optical depth is enough to significantly bias the power spectrum and the resulting posteriors, even if the temperature equation is solved with the local density value. 

    Since a part of the difference between the results of each code remains unexplained, and because they both naturally rely on modelling choices of uncertain accuracy we cannot claim that one is more accurate than the other. {We made the natural choice to use only the X-ray luminosity of the \textsc{Licorice} sources to compress them into a limited number of profiles, but an improved method to transfer \textsc{Licorice} sources to \textsc{Beorn} may lead to a better agreement of their 21-cm signals. The comparison between the unaltered \textsc{Licorice} power spectra and the power spectra of \textsc{Licorice} modified to incorporate Beorn-like approximations strongly suggests that it will be difficult to reach a better agreement than $\sim 20-30\%$ at the power spectrum level, and harder still to reach a sub-noise level of disagreement.}
    
    Also, the $~10\%$ difference observed in Figure \ref{fig:manyTk} cannot by itself explain the whole divergence at the power spectrum and inference stages: the mean temperatures are very close for the fiducial model and the inferences are still significantly different, while the posterior over $f_X$ obtained with the \textsc{Beorn} version of the spectrum is centred on a value of $f_X$ that is more than $10\%$ lower than the truth.   }
    { It seems likely that the codes studied here are similar enough for current observations by SKA pathfinders, but that considerable work is needed to make them ready for the SKA era.} {Still, we believe that an agreement of order $\sim 20 \%$ between their power spectra is a very good result given the complexity of the numerical modelling of the 21-cm signal, and this outcome is similar to that of previous comparison works such as \cite{Ghara2018} and \cite{Hutter2018} (which focused on ionisation rather than X-ray heating). }
    
    It is difficult to generalise these results to other codes such as \textsc{21cmFAST}, \textsc{Grizzly}, or C2ray, without explicitly including them in a similar comparison, which we leave for future work. However, {this work is intended as an attempt to test various modelling choices that are common in 21-cm simulation codes and } we already observe a significant difference between \textsc{Licorice} power spectra computed, for instance, with and without {a homogeneous optical depth}. This motivates a conservative claim in that this approximation is likely to cause significant disagreement between posteriors inferred using codes that use it and those that do not {\citep[such as \textsc{21cmFAST} for example, see][]{Mesinger2010}}. {More generally, we expect the scatter of Figure \ref{fig:inference_both} to be broadly representative of the impact of the approximations that we considered and thus of the result one might find when interpreting real data with several simulation codes. }

\section*{Acknowledgements}

The authors thank Aurel Schneider and Sambit Giri for the productive discussions and advice. JRP and RM acknowledge support from STFC grant ST/Y004132/1. 
This work was granted access to the HPC resources of MesoPSL financed
by the Region Ile de France and the project Equip@Meso (reference
ANR-10-EQPX-29-01) of the programme Investissements d’Avenir supervised
by the Agence Nationale pour la Recherche. 
This project was provided with computer and storage re-
sources by GENCI at TGCC thanks to the grant 2023-A0150413759 on the supercomputer Joliot Curie’s ROME partition.

\section*{Data Availability}

The data underlying this article will be shared on reasonable request to the corresponding author.



\bibliographystyle{mnras}
\bibliography{mnras_ads_clean} 




\newpage

\appendix

\section{Alternative source tracking and related issues}\label{sec:app_track}

Initially, the method used throughout this project to transfer \textsc{Licorice} sources to \textsc{Beorn} relied on the natural choice not to keep zero-luminosity `unborn' sources and to compute the average luminosity histories only on luminous sources. This led to a good agreement between the luminosity distributions of both versions of a given box.  

However, this meant that the number of sources belonging to a given luminosity bin was time-dependent. This forced \textsc{Beorn} to violate the conservation of energy and caused $\sim30 \% $ differences between the mean temperature  of \textsc{Licorice} and its Beorn-processed counterpart, although the emitted and deposited X-ray powers were identical in both versions.

The reason for this is that the correct total accumulated X-ray heating for bins $i$ containing $N_i(z)$ sources is, following Equation \ref{eq:rho_heat}:
\begin{multline}
       \langle T(z_{out}) \rangle   = T(z_{start}) + \\
     \frac{2}{3}\int_{z_{start}}^{z_{out}} \left[   \frac{3}{(1+z)}  \langle T(z) \rangle  - \frac{1}{k_B (1+z) H(z)}  \sum_i  \int_V [ N_i(z) \rho_X  ] \right]dz.
\end{multline}

The temperature, when computed this way in \textsc{Beorn} after including the cooling terms, does match the temperature in \textsc{Licorice}. However, what is being painted in \textsc{Beorn} is:
\begin{equation}
     \langle T(z_{out}) \rangle_{paint}   = T(z_{start}) +  \sum_i N_i(z_{out}) \int_{z_{start}}^{z_{out}} \int_V \frac{d\rho_T}{dz} dz,
\end{equation}
and the two approaches can only be equal if $N_i$ is in fact independent of $z$. {In other words, when the number of sources varies with time, painting temperature profiles around the sources that exist at a given redshift z as if they had always existed leads to biases in the mean temperature.}  

Hence, to preserve energy conservation we chose to include "unborn" sources of luminosity zero in the catalogues when the luminosity histories are constructed. Furthermore, to keep the  source population processed by \textsc{Beorn} close to the initial \textsc{Licorice} one, we do not paint profiles around those sources.

\section{Emitted and deposited power}\label{app:dep_power}

\begin{figure}
    \centering
    \includegraphics[scale = 0.5]{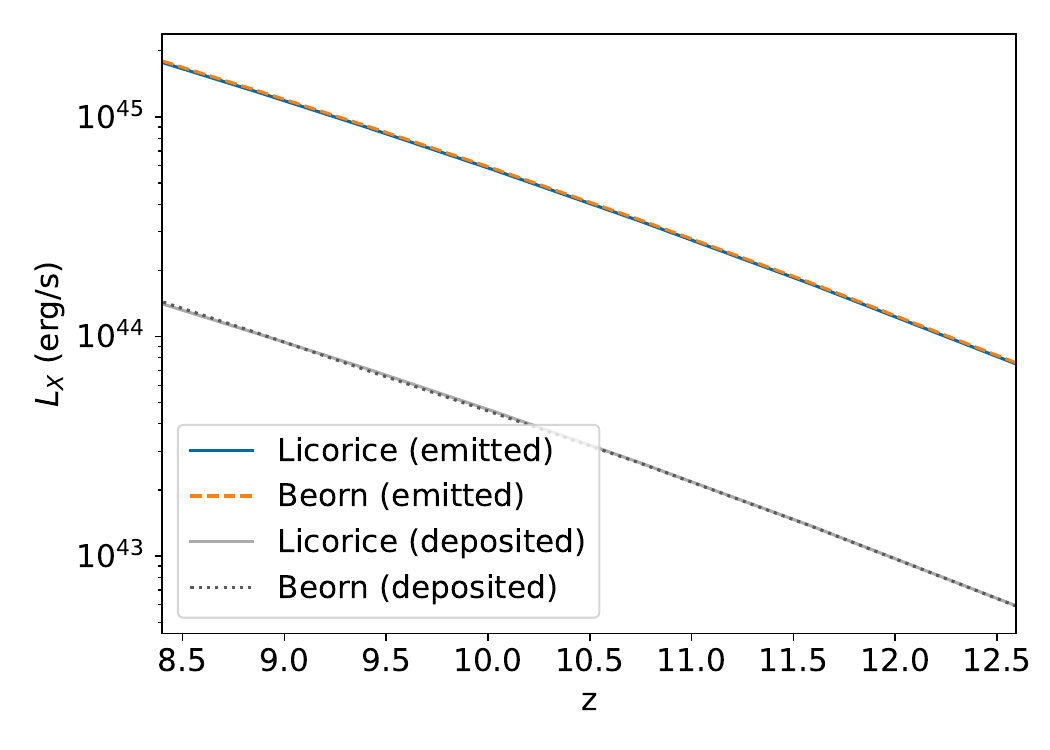}
    \caption{Total emitted and deposited power in the \textsc{Beorn} and \textsc{Licorice} version of the fiducial model.}
    \label{fig:lumX_eta}
\end{figure}

As a complement to Figure \ref{fig:lumX}, Figure \ref{fig:lumX_eta} shows both the emitted and deposited X-ray power of the fiducial model in the idealised setup. The emitted power is the luminosity of the sources measured in the box for both codes, while measuring of the deposited power was done in each code independently. In Licorice, it is simply derived by adding all the energy deposited by each photon packet in each cell during  radiative transfer. In Beorn, it is obtained by adding the integrated X-ray flux profiles $\rho_X$ during the painting step. Note that, as explained in the main text, these profiles are used by the temperature evolution to compute the temperature profiles.
In the idealised setup, since X-rays can deposit only a fraction $f_{heat} = 0.11$ of their energy,  $f_{heat}$ acts as an upper limit to the ratio of the emitted and deposited energy at a given time.  

This figure shows that the emitted and deposited energy agree nearly perfectly, and respect the $f_{heat}$ limit, as the "deposited-to-emitted" ratio is nearly constant at $\sim0.08$. This result is  consistent across the different models we have considered, and indicates that the source of the $\sim10\%$ mean temperature difference observed in Figure \ref{fig:manyTk} is likely  downstream of the X-ray propagation.

\section{temperature and 21-cm signal maps}\label{app:slices}

\begin{figure*}
    \centering
    \includegraphics[scale = 0.5]{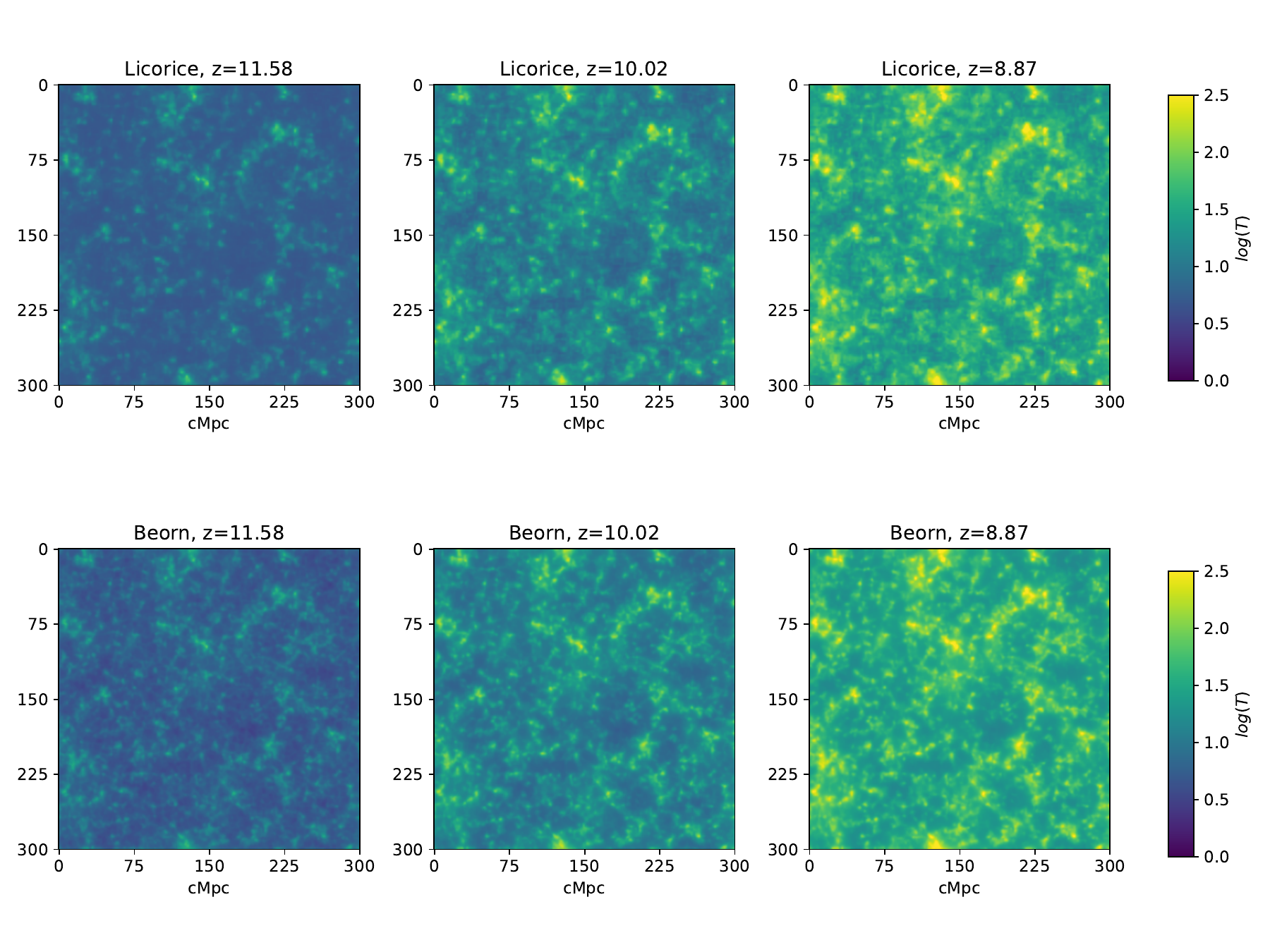}
        \includegraphics[scale = 0.5]{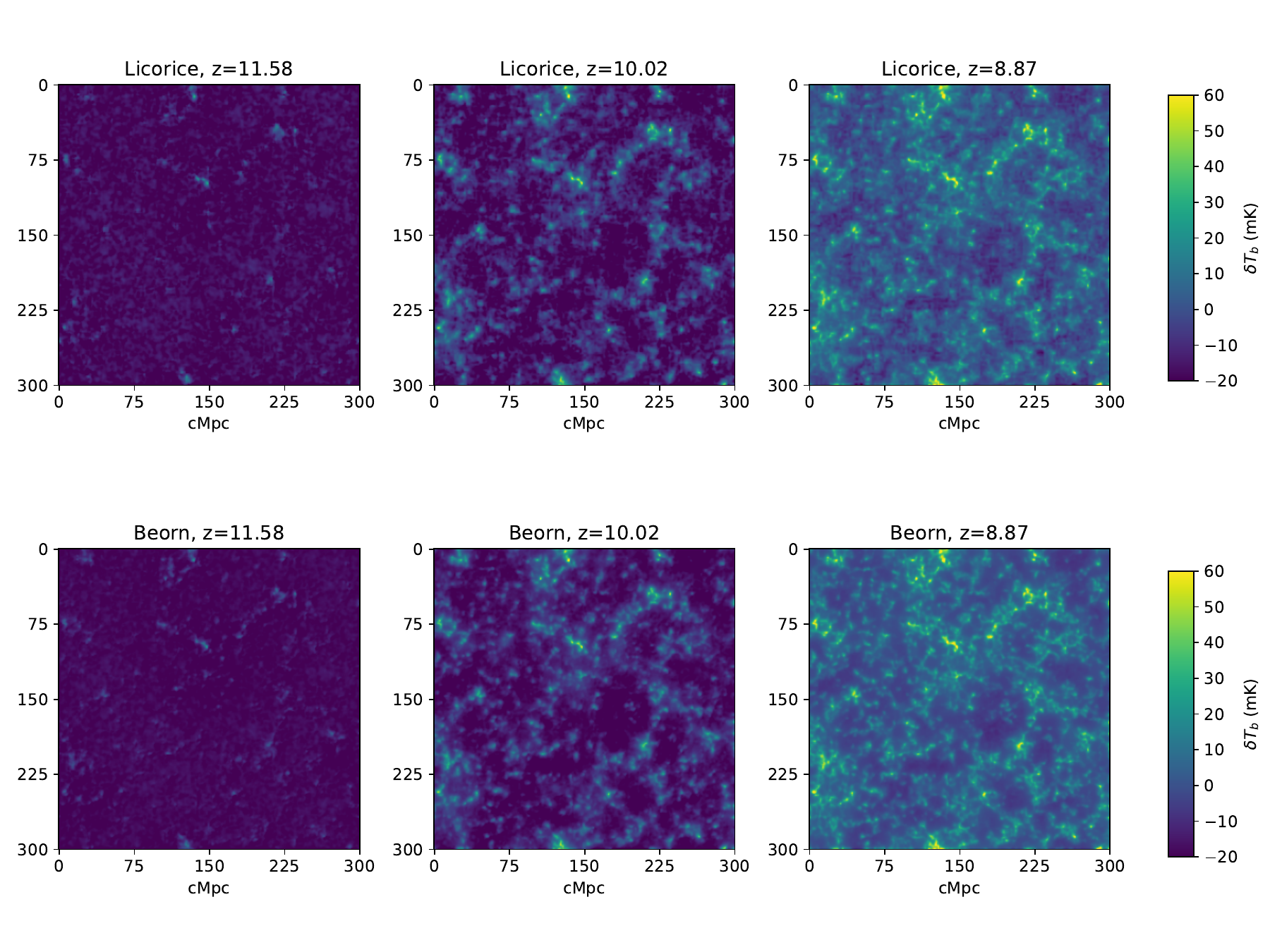}
     \caption{ \textit{First row}: temperature slices of the idealised version of fiducial \textsc{Licorice} simulation at redshifts $11.58, 10.02, 8.
     87$. \textit{Second row}: Same slices of the corresponding \textsc{Beorn} simulation. \textit{Third and fourth rows}: Same as the previous rows for the 21-cm brightness temperature. }
    \label{fig:Tkmaps_cstfxh}
\end{figure*}

\begin{figure*}
    \centering
    \includegraphics[scale = 0.5]{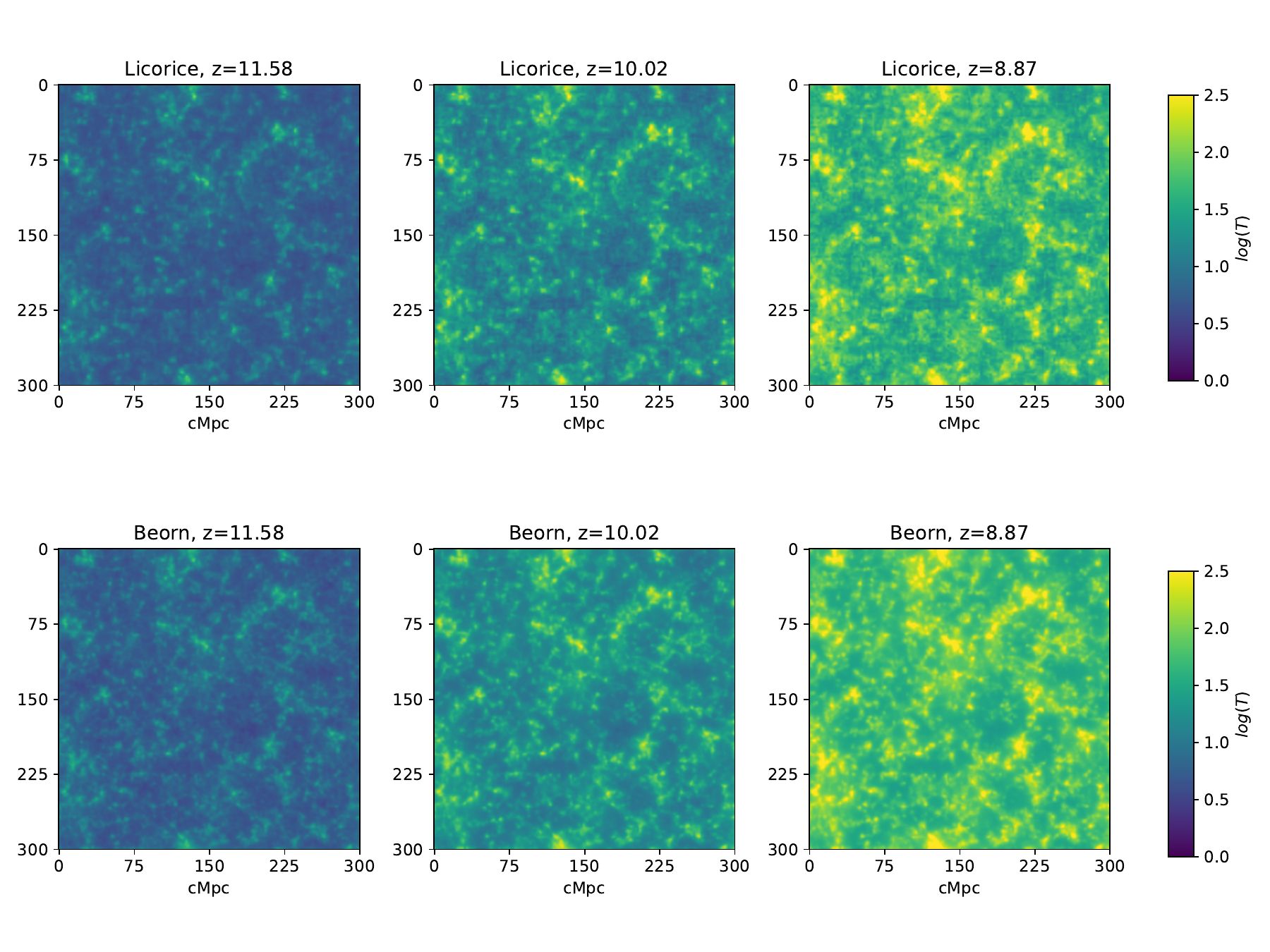}
        \includegraphics[scale = 0.5]{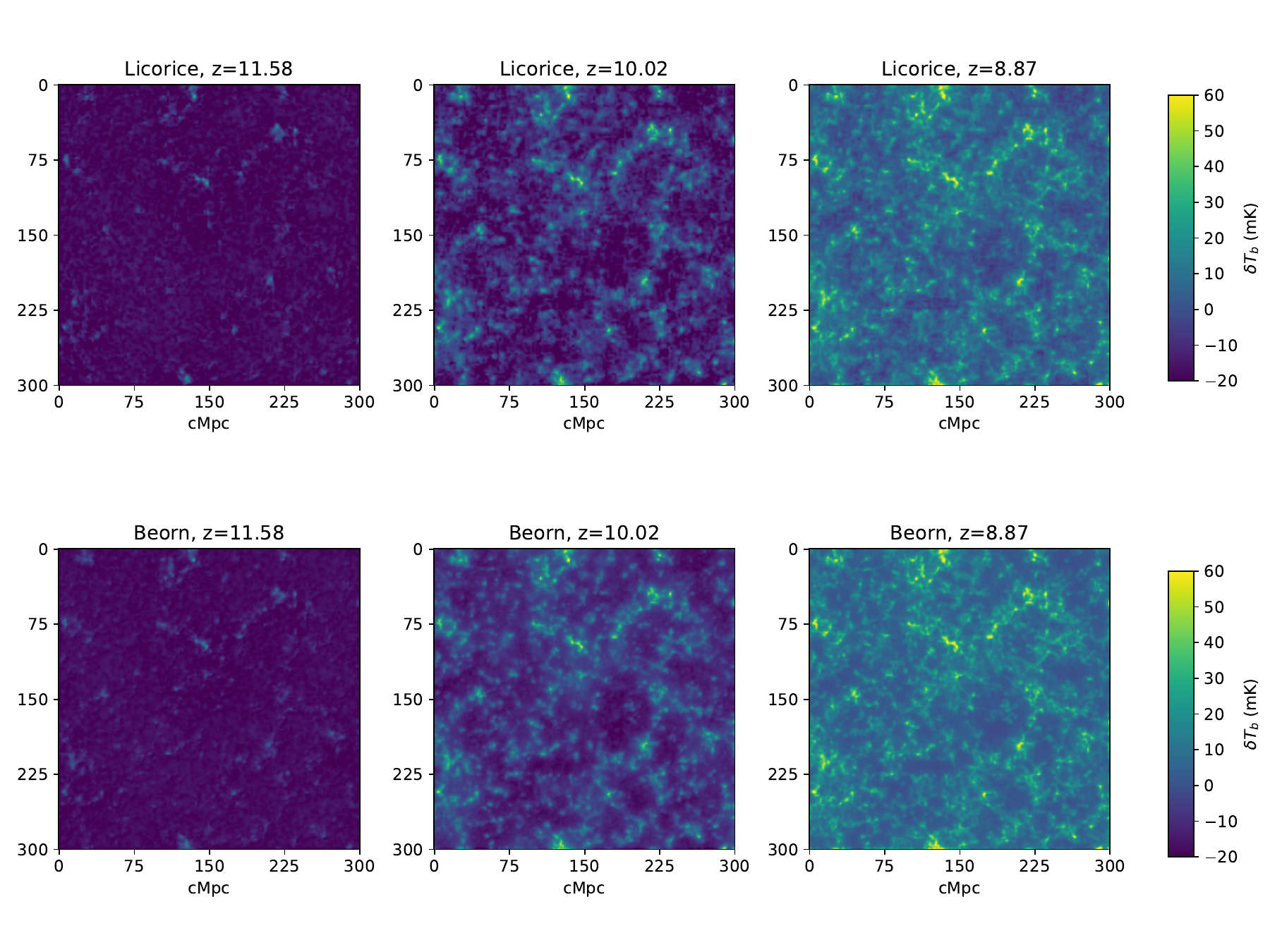}
    \caption{Same as Figure \ref{fig:Tkmaps_cstfxh}, but for the realistic version of the simulations.}
    \label{fig:Tkmaps}
\end{figure*}

{ Here we show in Figure \ref{fig:Tkmaps_cstfxh} and \ref{fig:Tkmaps} slices of the temperature cubes and 21-cm cubes of the \textsc{Beorn} and \textsc{Licorice} versions of the fiducial model, first in the idealised case, then in the realistic one. }These show visually the close agreement between the simulations in broad outline, although differences in the details can be seen on close examination.

\section{Dependence on the number of profile bins}

\begin{figure*}
    \centering
\includegraphics[scale=0.35]{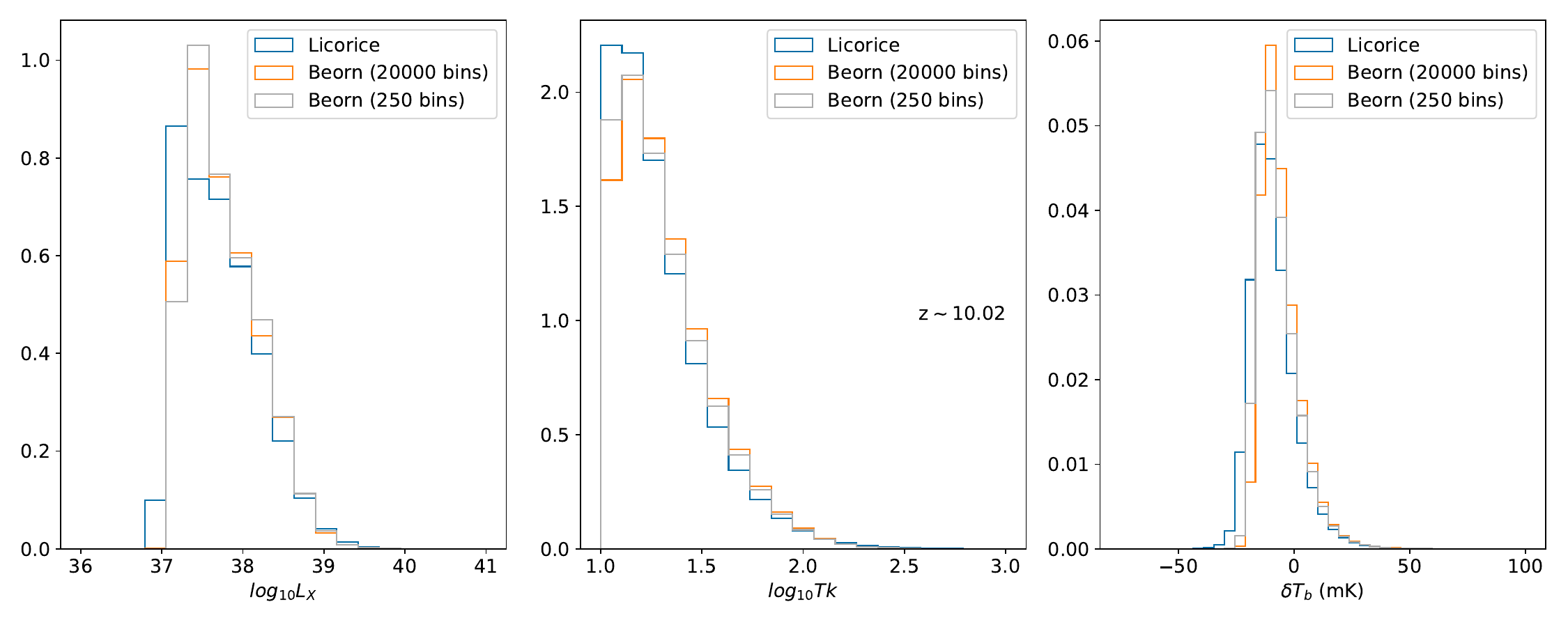}
    \caption{Evolution of the luminosity (\textit{left}), temperature (\textit{middle}), and 21-cm brightness temperature (\textit{right}) distributions with the number of luminosity bins (and therefore temperature profiles) computed by \textsc{Beorn}. The distributions of the corresponding \textsc{Licorice} box are also plotted. }   \label{fig:meanq_nbins}
\end{figure*}

\begin{figure}
    \centering
    \includegraphics[scale=0.5]{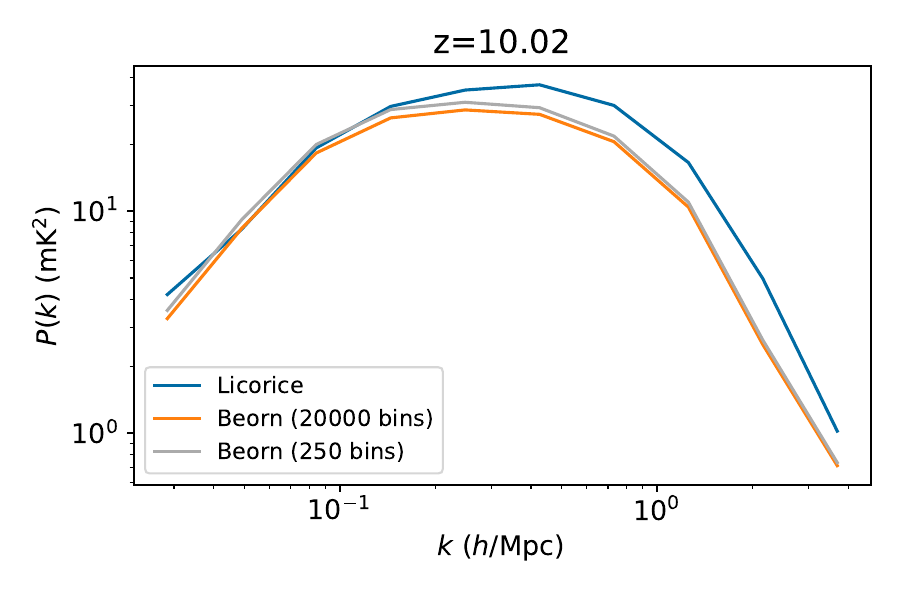}
    \caption{Evolution of the power spectrum at $z=10$ as the number of profile bins increases. The \textsc{Licorice} spectrum of the corresponding box is also plotted. }
    \label{fig:pk_nbins}
\end{figure}

\textsc{Beorn} computes a finite number of temperature profiles that are then painted around the corresponding sources. As explained in Section \ref{sec:bin_proc}, we define $N$ luminosity bins. In this appendix, we show how relevant physical quantities evolve with $N$. To do so, we run \textsc{Beorn} simulations on \textsc{Licorice} sources (model 1)  with $N=250$ and $N=20000$. Note that the computation cost of \textsc{Beorn} scales linearly with $N$. The $N=250$ case runs in a few cpu hours, and $N=20000$ in a few hundred cpu hours, about as much as \textsc{Licorice} at $256^3$ resolution.

First, Figure \ref{fig:meanq_nbins} shows the mean luminosity, temperature, and brightness temperature at $z=10$.
Between $N=250$ and $N=20000$ the difference
in mean luminosity is $\sim3\%$, in mean temperature $\sim5\%$, and in mean 21-cm signal $\sim10\%$. The latter two are small compared to the difference between \textsc{Beorn} and \textsc{Licorice}. Figure \ref{fig:pk_nbins} shows the 21-cm power spectra of those simulations. Again, the effect of using 20000 profiles instead of 250 is $\sim 15\%$. This is not negligible compared to what is shown in Figure \ref{fig:powerspectrum}, but not  the full effect. It is possible that using millions of profiles improves the agreement between the power spectra, but this would render \textsc{Beorn} extremely expensive.

\section{Single-source toy model}

We have run \textsc{Licorice} and \textsc{Beorn} in the idealised setup, but forced \textsc{Licorice} to form only a single source near the centre of the simulated box. The goal is to measure the temperature profile around the source. We also limit the X-ray spectrum to the [100,110]
eV range in order for most of the heating to occur near the source.

\begin{figure}
    \centering
\includegraphics[scale=0.6]{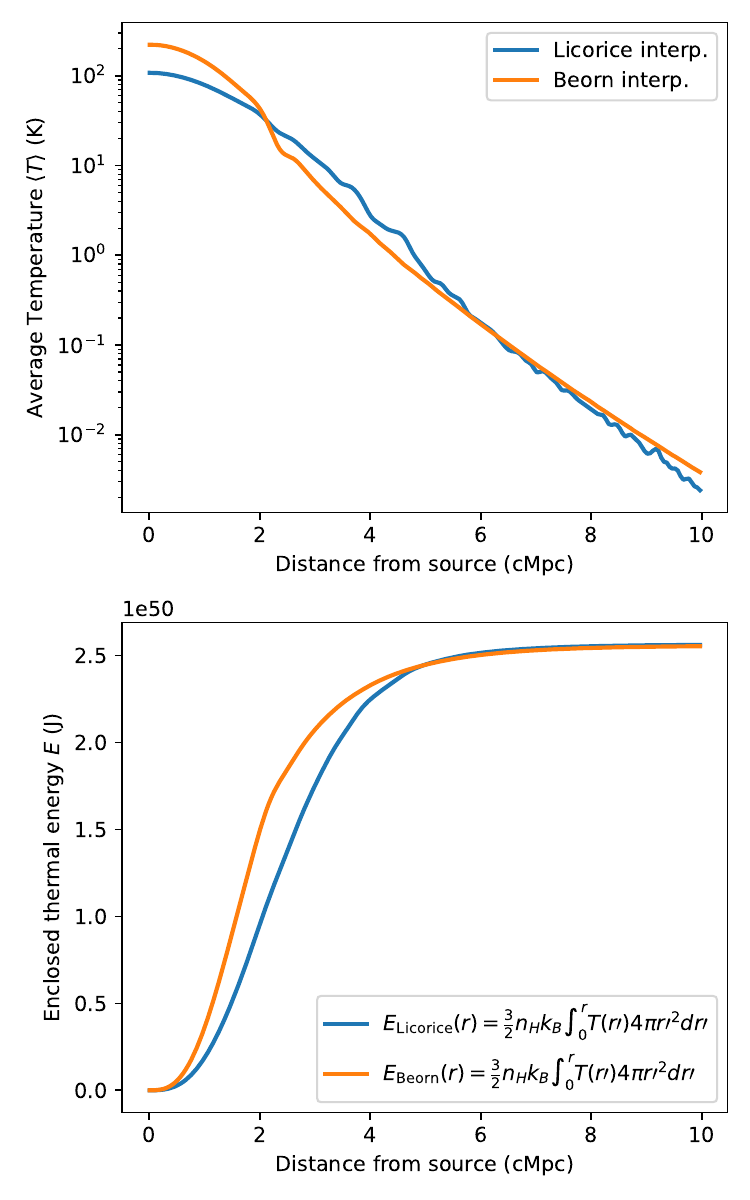}
    \caption{\textit{Top:} temperature profile around the single source in the box. \textit{Bottom:} Thermal energy enclosed in the sphere of a given radius around the single source.  }   \label{fig:toymodel_profiles}
\end{figure}

Figure \ref{fig:toymodel_profiles} shows the radial temperature profile around the source as modelled by \textsc{Beorn} and \textsc{Licorice} at $z\approx6.2$, as well as the thermal energy enclosed in the sphere centred around the source. The adiabatically cooled background has been subtracted, so that only the excess heating by the source is considered here. We observe that the total heating is extremely similar in \textsc{Licorice} and \textsc{Beorn}. We also observe that the profiles have a slightly different shape, with \textsc{Beorn} having a higher peak near the source, while \textsc{Licorice} shows a smoother radial evolution. We hypothesise that this shape difference could be explained by the particles-to-RT-cells-to-particles process that occurs during the \textsc{Licorice} simulation and is followed by a final arrangement of the particles on a grid to produce e.g.the temperature, density, ionisation cubes. This process should however conserve energy.

\section{Temperature power spectra}\label{sec:app_t_pk}

\begin{figure}
    \centering
    \includegraphics[scale = 0.55]{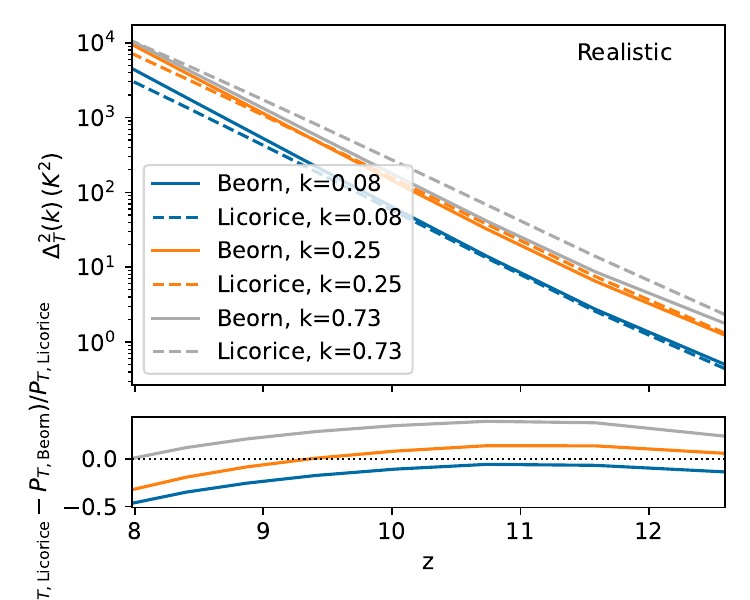}
    \includegraphics[scale = 0.55]{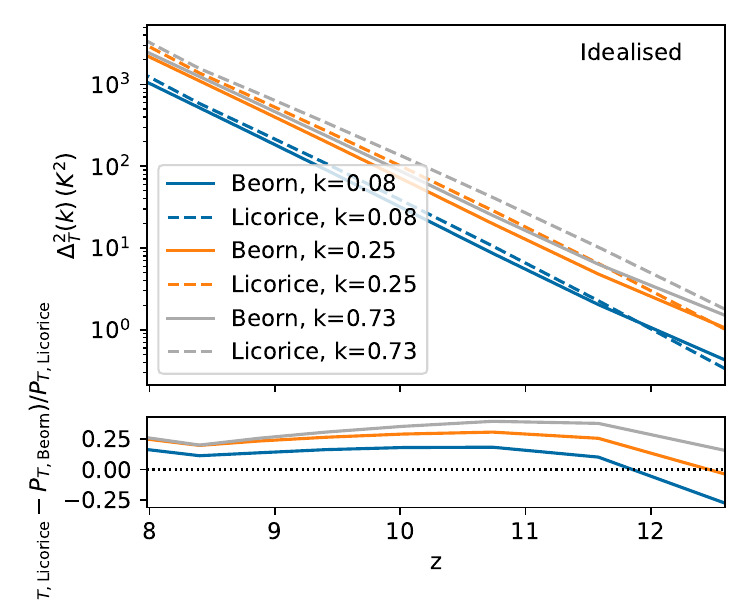}
    \caption{Similar to Figure \ref{fig:powerspectrum}, but for power spectra of the temperature fields instead.}
    \label{fig:T_powerspectrum}
\end{figure}

{We show on Figure \ref{fig:T_powerspectrum} the power spectra of the temperature fields. Although the relative difference between Beorn and Licorice is not exactly the same as on Figure \ref{fig:powerspectrum}, it shows that the distribution of heating is different by $\sim 25\%$. This difference will also affect the correlations with the density and Lyman-$\alpha$ fields, which in turn result in Figure \ref{fig:powerspectrum}. The relative differences are not the same in the realistic and idealised setups and are non-negligible in the latter, which suggests that unidentified effects remain, possibly caused by the preprocessing of the sources.  }

\bsp	
\label{lastpage}
\end{document}